\documentclass[aps,prd,preprint,superscriptaddress]{revtex4-2}
\usepackage{ulem}
\usepackage{xcolor}
\usepackage[colorlinks=true]{hyperref}
\usepackage{soul}
\usepackage{booktabs}

\usepackage{amsmath,amssymb,color,epsfig}
\allowdisplaybreaks[4]

\newcommand{\be}{\begin{equation}}
\newcommand{\ee}{\end{equation}}
\newcommand{\bea}{\setlength\arraycolsep{2pt} \begin{eqnarray}}
\newcommand{\eea}{\end{eqnarray}}
\newcommand{\nn}{\nonumber}

\begin{document}

\hypersetup{
    linkcolor=blue,
    citecolor=red,
    urlcolor=magenta
}

%\preprint{}

\title{Neutron stars more compact than black holes in quasi-topological gravity: Equilibrium configurations and radial stability}
%Stability of radial oscillations of neutron stars more compact than black holes



\author{Liang Liang}
\affiliation{Department of Physics, Key Laboratory of Low Dimensional Quantum Structures and Quantum Control of Ministry of Education, and Institute of Interdisciplinary Studies, Hunan Normal University, Changsha, 410081, China}
\affiliation{Hunan Research Center of the Basic Discipline for Quantum Effects and Quantum Technologies, Hunan Normal University, Changsha 410081, China}

\author{Zhe Luo}
\affiliation{Department of Physics, Key Laboratory of Low Dimensional Quantum Structures and Quantum Control of Ministry of Education, and Institute of Interdisciplinary Studies, Hunan Normal University, Changsha, 410081, China}
\affiliation{Hunan Research Center of the Basic Discipline for Quantum Effects and Quantum Technologies, Hunan Normal University, Changsha 410081, China}

\author{Shoulong Li}
\email[Corresponding author: ]{shoulongli@hunnu.edu.cn}
\affiliation{Department of Physics, Key Laboratory of Low Dimensional Quantum Structures and Quantum Control of Ministry of Education, and Institute of Interdisciplinary Studies, Hunan Normal University, Changsha, 410081, China}
\affiliation{Hunan Research Center of the Basic Discipline for Quantum Effects and Quantum Technologies, Hunan Normal University, Changsha 410081, China}

\author{Hongwei Yu}
\email[]{hwyu@hunnu.edu.cn}
\affiliation{Department of Physics, Key Laboratory of Low Dimensional Quantum Structures and Quantum Control of Ministry of Education, and Institute of Interdisciplinary Studies, Hunan Normal University, Changsha, 410081, China}
\affiliation{Hunan Research Center of the Basic Discipline for Quantum Effects and Quantum Technologies, Hunan Normal University, Changsha 410081, China}


\date{\today}

\begin{abstract}

Within general relativity, black holes are widely regarded as the ultimate benchmark for compactness in the Universe. Recently, however, neutron star models have been constructed in a higher-curvature theory---quasi-topological gravity (QTG)---whose compactness can exceed the black-hole limit~ [S. Li, H. L\"u, Y. Gao, R. Xu, L. Shao, and H. Yu, companion Letter, Neutron stars more compact than black holes as a probe of strong-field gravity, Phys. Rev. D 114, L021504 (2026).]. Here we present a detailed analysis of both the equilibrium structure and radial stability of such configurations in QTG. By examining several representative equations of state and different values of the gravitational coupling constant, we find that in the high-central-density regime the compactness exceeding the black-hole bound exhibits a universal behavior in QTG. We further show that QTG corrections grow increasingly significant at large central densities and can stabilize configurations that are radially unstable in general relativity over a broad parameter range. These results establish ultra-compact neutron stars in QTG as theoretically viable strong-field configurations and provide a foundation for further investigations of their dynamical and phenomenological implications.

\end{abstract}
%\keywords{}

\maketitle

\section{Introduction}

Over the past century, general relativity (GR) has successfully passed all precision tests in weak-field regimes and many tests in strong-field regimes~\cite{Will:2014kxa,Berti:2015itd,Barack:2018yly,Freire:2024adf,Psaltis:2008bb}. Its most profound predictions have been further confirmed by direct observations, including the detection of gravitational waves from compact binary coalescences~\cite{LIGOScientific:2016aoc} and the imaging of photon rings around supermassive black holes~\cite{EventHorizonTelescope:2019dse}. Together, these achievements firmly establish GR as a cornerstone of modern fundamental physics. Despite its empirical success, GR is fundamentally incompatible with quantum mechanics. From a field-theoretic perspective, GR is neither ultraviolet complete nor perturbatively renormalizable and should therefore be regarded as an effective field theory valid below a certain energy scale~\cite{Stelle:1976gc,Stelle:1977ry,Donoghue:1994dn}. This viewpoint naturally suggests that GR may receive corrections at high energies or in extremely strong gravitational fields. A systematic way to parameterize such deviations is through higher-curvature terms, which arise generically in effective field theory treatments of gravity and also appear in candidate theories of quantum gravity such as string theory~\cite{Callan:1985ia}. Importantly, higher-curvature corrections are not merely of theoretical interest; they can have concrete phenomenological implications. A notable example is Starobinsky gravity~\cite{Starobinsky:1980te}, in which a simple quadratic curvature term drives a successful inflationary scenario, demonstrating that higher-curvature effects can play a physically meaningful role.

The inclusion of higher-curvature terms generically modifies the Einstein field equations by introducing higher-order derivatives. As a result, the dynamical structure of the theory can differ substantially from that of GR, whose field equations are second order. This distinction is particularly transparent in the context of gravitational collapse. In GR~\cite{Misner:1973prb,Poisson:2009pwt}, assuming that the spacetime satisfies the Einstein equations, the matter content obeys the weak and strong energy conditions, and a trapped surface forms during evolution, the Penrose singularity theorem~\cite{Penrose:1964wq} implies that the spacetime must be geodesically incomplete, signaling the formation of a black hole. In spherical symmetry, Birkhoff's theorem further guarantees that the exterior spacetime remains Schwarzschild throughout the collapse. Consequently, once the stellar radius contracts below the Schwarzschild radius, the formation of a trapped surface and the subsequent emergence of a black hole are unavoidable.

In contrast, gravitational collapse in higher-curvature gravity theories constitutes a genuinely higher-order initial value problem. Due to the violation of Birkhoff's theorem, even in spherical symmetry the exterior spacetime is no longer constrained to remain Schwarzschild during evolution. As a result, the formation of trapped surfaces is no longer automatic, even when the matter sector continues to satisfy the weak and strong energy conditions. From the perspective of initial data, higher-curvature terms modify the region of phase space that leads to trapped surface formation, rendering gravitational collapse more subtle than in GR. While certain initial conditions may still result in black hole formation, other admissible initial data can instead give rise to horizonless configurations that are even more compact than black holes and remain dynamically stable over astrophysically relevant timescales.

Although a fully dynamical treatment of gravitational collapse in higher-curvature gravity is technically challenging, this dynamical picture naturally suggests the possibility of stellar remnants that are more compact than black holes. Since the most compact known stellar objects in the Universe are neutron stars, whose structure is supported predominantly by nuclear interactions, it is natural to expect that any such ultra-compact, horizonless objects---if realized in nature---are likely composed of neutron star matter or perhaps even more exotic phases of matter at higher densities. Framing them in this way ensures that these theoretical configurations remain consistent with existing astrophysical observations and phenomenological constraints.

Along this line, the concept of neutron star configurations whose compactness exceeds that of black holes was explicitly proposed in Ref.~\cite{Li:2023vbo}, where explicit models were constructed within the framework of quasi-topological gravity (QTG)~\cite{Li:2017ncu,Hennigar:2017ego}. QTG provides an ideal theoretical framework for studying such ultra-compact stellar objects. First, it admits a richer class of spherically symmetric vacuum solutions than GR, reflecting the breakdown of Birkhoff’s theorem, while the Schwarzschild metric remains the unique spherically symmetric black hole solution of the theory. This enlarged vacuum structure allows the exterior spacetime of a compact star to differ from that of a black hole, thereby permitting horizonless stellar configurations with radii smaller than the Schwarzschild radius. Second, QTG exactly reduces to GR in the weak-field regime~\cite{Li:2017ncu,Hennigar:2017ego}, ensuring consistency with current observational constraints. At the same time, it avoids introducing additional propagating degrees of freedom in the linearized vacuum spectrum, such as ghost-like massive spin-2 modes~\cite{Stelle:1977ry}.

Reference~\cite{Li:2023vbo} constructed neutron star models that can become more compact than black holes and provided a preliminary assessment of their radial stability for a representative stellar configuration, thereby leaving open two important questions: whether this behavior---namely, the emergence of neutron stars exceeding the black-hole compactness threshold in the high-central-density regime for specific values of the gravitational coupling constant and a given equation of state (EOS)---is a generic feature that persists for other values of the coupling constant and alternative EOSs, and how the radial stability of these ultra-compact configurations depends on the gravitational coupling parameter. In the present work, we address these issues through a comprehensive investigation of both the equilibrium structure and the radial stability of these configurations.

The remainder of this paper is organized as follows. In Sec.~\ref{framework}, we briefly review the formulation of cubic QTG and construct the corresponding black hole solutions, establishing the Schwarzschild solution as the unique black hole and a benchmark for compactness. In Sec.~\ref{NSinQTG}, we derive the modified Tolman-Oppenheimer-Volkoff (TOV) equations and study the neutron star equilibrium configurations. In Sec.~\ref{Stability}, we analyze the radial perturbation equations and carry out a detailed investigation of radial oscillations and stability. Finally, Sec.~\ref{Conclusion} summarizes our main results and discusses their implications.

\section{Quasi-topological Ricci cubic gravity} \label{framework}

\subsection{The theory and its equation of motion}

The QTG model considered in this work is constructed from cubic polynomials of the Ricci tensor. It represents the simplest nontrivial realization within the broader class of QTG theories~\cite{Li:2017ncu,Hennigar:2017ego}. The action $S$ takes the form
\be
S = \frac{c^4}{16 \pi G} \int{ d^4 x \sqrt{-g} L} +S_\textup{m} \,, \label{action}
\ee
where $g$ denotes the determinant of the spacetime metric $g_{\mu\nu}$, $L$ is the gravitational Lagrangian density, and $S_{\textup{m}}$ represents the matter action. The symbols $c$ and $G$, representing the speed of light and the gravitational constant, are set to unity ($c=G=1$) in the remainder of the work, as we adopt geometric units. The gravitational Lagrangian of the QTG model is given by~\cite{Li:2017ncu,Hennigar:2017ego}
\begin{equation}
L= {\cal R} + \lambda ({\cal R}^3-6{\cal R}{\cal R}_{\mu\nu}{\cal R}^{\mu\nu}+8{{\cal R}_\mu}^\nu{{\cal R}_\nu}^\gamma{{\cal R}_\gamma}^\mu) \,,  \label{qtg}
\end{equation}
where ${\cal R}$ and ${\cal R}_{\mu\nu}$ denote the Ricci scalar and Ricci tensor, respectively, and $\lambda$ is the coupling constant associated with the quasi-topological combination. The indices are raised and lowered using the metric $g_{\mu\nu}$ and its inverse $g^{\mu\nu}$. Varying the action~(\ref{action}) with respect to the metric yields the equation of motion (EOM) for gravitational field:
\begin{equation}
E_{\mu\nu} \equiv {\cal P}_{\mu\alpha\beta\gamma}{{\cal R}_\nu}^{\alpha\beta\gamma}-\frac{1}{2}g_{\mu\nu}L-2\nabla^\alpha\nabla^\beta {\cal P}_{\mu\alpha\beta\nu} =  8\pi T_{\mu\nu} \,, \label{eom}
\end{equation}
where ${\cal R}_{\mu\alpha\beta\gamma}$ is the Riemann tensor, $\nabla_\mu$ denotes the covariant derivative, and the tensor ${\cal P}_{\mu\alpha\beta\gamma}$ is defined as
\be
{\cal P}_{\mu\alpha\beta\gamma} \equiv \frac{\partial L}{\partial {\cal R}^{\mu\alpha\beta\gamma}} \,.
\ee
The energy-momentum tensor $T_{\mu\nu}$ of matter is given by
\be
T_{\mu\nu} \equiv -\frac{2}{\sqrt{-g}} \frac{\partial S_\textup{m}}{\partial g^{\mu\nu}} \,.
\ee
For each curvature invariants present in the Lagrangian~(\ref{qtg})
\be
 \{{\cal R},\ {\cal R}^3,\ {\cal R}{\cal R}_{\mu\nu}{\cal R}^{\mu\nu},\ {{\cal R}_\mu}^\nu{{\cal R}_\nu}^\gamma{{\cal R}_\gamma}^\mu \}, \nn
\ee
the constituent components of the  ${\cal P}$-tensor are derived as follows:
\bea
{\cal P}^{(1)}_{\mu\alpha\beta\gamma} &= & \frac{1}{2} (g_{\mu\beta}g_{\alpha\gamma} -g_{\mu\gamma}g_{\alpha\beta} ) \,, \nn \\
{\cal P}^{(2)}_{\mu\alpha\beta\gamma} &= & \frac{3}{2} {\cal R}^2 (g_{\mu\beta}g_{\alpha\gamma} -g_{\mu\gamma}g_{\alpha\beta} ) \,, \nn \\
{\cal P}^{(3)}_{\mu\alpha\beta\gamma} &= & \frac{1}{2} {\cal R}_{\rho\sigma}  {\cal R}^{\rho\sigma} (g_{\mu\beta}g_{\alpha\gamma} -g_{\mu\gamma}g_{\alpha\beta} ) +\frac{1}{2} {\cal R} ( g_{\mu\beta} {\cal R}_{\alpha\gamma} -g_{\mu\gamma} {\cal R}_{\alpha\beta} -g_{\alpha\beta} {\cal R}_{\mu\gamma} +g_{\alpha\gamma} {\cal R}_{\mu\beta})   \,, \nn \\
{\cal P}^{(4)}_{\mu\alpha\beta\gamma} &= & \frac{3}{4} (g_{\mu\beta} {\cal R}_{\alpha\nu}  {\cal R}_{\gamma}{}^{\nu} -g_{\mu\gamma} {\cal R}_{\alpha\nu}  {\cal R}_{\beta}{}^{\nu} - g_{\alpha\beta} {\cal R}_{\mu\nu}  {\cal R}_{\gamma}{}^{\nu} +g_{\alpha\gamma} {\cal R}_{\mu\nu}  {\cal R}_{\beta}{}^{\nu}) \,. 
\eea

\subsection{Black holes in quasi-topological gravity}

We now investigate static and spherically symmetric black hole solutions in the QTG model~(\ref{qtg}) and address their uniqueness. Setting the energy-momentum tensor to zero in Eq.~(\ref{eom}), it follows immediately that the theory admits Einstein metrics as vacuum solutions. In particular, all Ricci-flat solutions of GR, satisfying ${\cal R}_{\mu\nu}=0$, automatically solve the QTG field equations. This ensures that the Schwarzschild spacetime is a valid vacuum solution of the theory. However, the existence of the Schwarzschild solution alone does not guarantee its uniqueness. In order to determine whether it represents the unique static and spherically symmetric black hole configuration in QTG, a more detailed analysis is required.

We consider the most general static and spherically symmetric line element,
\be
ds^2 = -h dt^2+f^{-1} dr^2+r^2(d\theta^2+\sin^2\theta d\phi^2)  \,, \label{ansatz}
\ee
where $h$ and $f$ are functions of the radial coordinate $r$. Substituting this ansatz into the field equations~(\ref{eom}), one finds that only diagonal components remain nonvanishing. The resulting system reduces to two independent ordinary differential equations (ODEs) for $h$ and $f$
\begin{widetext}
\begin{eqnarray}
&&12 \lambda r^3 f^2 h^2 (f h'-h f') h'''  -\Big[\lambda (-3 (h f'-f h') \big(2 f h (2 r^3 f' h'^2+h^2 (r f'+8)-r h h') \nonumber \\
&&+r h^2 f' (r^2 f' h'+2 h) +f^2 (-5 r^3 h'^3+6 r h^2 h'-16 h^3)\big) -6 r f h^2 f'' (-3 r^2 h f' h'  \nonumber\\
&&+3 r^2 f h'^2+4 (f-1) h^2)+h'' \big(12 r^3 f^2 h^3 f''+6 r f h (-f h (7 r^2 f' h'+4 h) +2 r^2 h^2 f'^2  \nonumber\\
&&+f^2 (5 r^2 h'^2+4 h^2))\big) -12 r^3 f^3 h^2 h''^2 ) +2 r^3 h^3 (r f h'+(f-1) h)\Big] =0 \,, \label{eom1}\\
\text{and} \nonumber\\
&& 12 \lambda r^3 h^3 f (f h'-h f')  f'''  -\Big[12 \lambda r^3 f h^4 f''^2-3 \lambda h^3 (r^3 f'^3-4 f^2 (5 r f'+4)+12 r f f' \nonumber\\
&&+16 f^3) h'-45 \lambda r^3 f^2 h f' h'^3+9 \lambda r f h^2 (3 r^2 f'^2-2 f^2+2 f) h'^2+2 h^4 (f (-21 \lambda r f'^2 \nonumber\\
&&-24 \lambda f'+r^3)+9 \lambda r f'^2+24 \lambda f^2 f'+r^4 f'-r^3)+6 \lambda r h^2 f'' (f h (4 h-5 r^2 f' h') \nonumber\\
&&+r^2 h^2 f'^2-4 f^2 (h^2-r^2 h'^2))+h'' (-12 \lambda r^3 f^2 h^3 f''-6 \lambda r f h (f h (4 h-5 r^2 f' h') \nonumber\\
&&+r^2 h^2 f'^2-4 f^2 (h^2-r^2 h'^2)))+21 \lambda r^3 f^3 h'^4 \Big] =0 \,, \label{eom2}
\end{eqnarray}
\end{widetext}
where a prime denotes the derivative with respect to $r$. Following the strategy of Ref.~\cite{Lu:2015cqa}, we assume the presence of an event horizon located at $r=r_0>0$, where both metric functions vanish. Near the horizon, $h$ and $f$ admit Taylor expansions of the form
\bea
\lim_{r\rightarrow r_0}h(r) &=& \sum_{n=1}^\infty h_n (r- r_0)^n \,, \nn \\
\lim_{r\rightarrow r_0}f(r) &=& \sum_{n=1}^\infty f_n (r- r_0)^n \,. \label{nearhorizon}
\eea
The coefficient $h_1$ remains free due to the freedom of rescaling the time coordinate and is fixed by imposing asymptotic flatness at spatial infinity. Substituting these expansions into Eqs.~(\ref{eom1})--(\ref{eom2}) yields two distinct and disconnected branches of near-horizon boundary conditions. The first branch coincides precisely with the Schwarzschild solution,
\begin{eqnarray}
&\textup{Case I}:&\quad  h_2 = -\frac{1}{r_0}  \,, \quad h_3 = \frac{1}{r_0^2} \,,  \quad \dots \,, \nonumber\\
&&\quad  f_1 = \frac{1}{r_0} \,,\quad f_2 = -\frac{1}{r_0^2}  \,, \quad f_3 = \frac{1}{r_0^3} , \quad \dots \,. \label{case1} 
\end{eqnarray}
The second branch represents an alternative branch where the coefficients depend on the coupling constant $\lambda$
\begin{eqnarray}
&\textup{Case II}:&\quad  h_2 = -\frac{1}{r_0} +\frac{r_0^3}{24\lambda} \,, \quad h_3 = \frac{1}{r_0^2} -\frac{3 r_0^2}{40\lambda} +\frac{r_0^6}{288\lambda^2} \,,  \quad \dots \,, \nonumber\\
&&\quad  f_1 = \frac{1}{r_0} \,,\quad f_2 = -\frac{1}{r_0^2} -\frac{r_0^2}{8\lambda} \,, \quad f_3 = \frac{1}{r_0^3} -\frac{r_0}{40\lambda} -\frac{r_0^5}{288\lambda^2} \,, \quad \dots \,. \label{case2}
\end{eqnarray}
Notably, for a fixed $\lambda$ and $r_0$, both branches are fully determined without additional shooting parameters. (The Taylor expansions are computed up to ${\cal O}((r - r_0)^{19})$.) Consequently, the solutions for Case~I~(\ref{case1}) and Case~II~(\ref{case2}) represent two bifurcating branches, with no integration constants interpolating between them. At large radial distances, asymptotic flatness requires that both metric functions admit expansions in inverse powers of $r$,
\bea
\lim_{r\rightarrow \infty}h(r) &=& \sum_{n=0}^\infty {h}^{\textup{inf}}_n r^{-n} \,, \nn \\
\lim_{r\rightarrow \infty}f(r) &=& \sum_{n=0}^\infty {f}^{\textup{inf}}_n r^{-n} \,, \label{nearinfinity}
\eea
Substituting Eq.~(\ref{nearinfinity}) into Eqs.~(\ref{eom1})--(\ref{eom2}), we obtain
\begin{equation}
\lim_{r\rightarrow\infty}h(r) = \lim_{r\rightarrow\infty}f(r) = 1 - \frac{2 M}{r} \,, \label{infinity}
\end{equation}
where $M$ is the gravitational mass. 

In generic higher-curvature gravity theories, linearization around flat spacetime typically reveals Yukawa-type corrections associated with additional massive scalar or massive spin-2 modes~\cite{Stelle:1977ry}. In contrast, the specific quasi-topological combination appearing in Eq.~(\ref{qtg}) is constructed such that no extra propagating degrees of freedom arise at the linearized level~\cite{Li:2017ncu,Hennigar:2017ego}. This property can be verified directly through a weak-field analysis. The weak-field vacuum spacetime of a static and spherically symmetric self-gravitating object can be treated as a linear perturbation around a maximally symmetric background, expressed as 
\be
g_{\mu\nu} = \bar{g}_{\mu\nu} + \tilde{g}_{\mu\nu} \,, 
\ee
where ``bar'' and ``tilde'' denote quantities associated with the background and perturbations, respectively.  For the maximally symmetric backgrounds, the Riemann tensor takes the form 
\be
\bar{{\cal R}}_{\mu\nu\rho\sigma} = K ( \bar{g}_{\mu\rho}\bar{g}_{\nu\sigma} - \bar{g}_{\mu\sigma}\bar{g}_{\nu\rho}) \,, 
\ee
with $K$ a constant. The linearized vacuum gravitational field equation in QTG is given by
\begin{equation}
 \tilde{E}_{\mu\nu} \equiv \delta \tilde{{\cal R}}_{\mu\nu} -\frac12 \bar{g}_{\mu\nu}  \delta \tilde{{\cal R}} =0 \,, \label{weak}
\end{equation} 
with 
\be
\delta \tilde{{\cal R}}_{\mu\nu} =  \tilde{{\cal R}}_{\mu\nu} - K \tilde{g}_{\mu\nu} \,, \quad  \textup{and} \quad   \delta \tilde{{\cal R}} = \bar{g}^{\mu\nu} \delta \tilde{{\cal R}}_{\mu\nu} \,. 
\ee
These equations coincide with the linearized Einstein equations in vacuum. Although the full vacuum field equations~(\ref{eom}) contain higher-derivative terms, they reduce to a second-order system in the weak-field regime. The solution therefore contains a single integration constant associated with the mass, and the spectrum consists only of the standard massless graviton, without additional scalar or ghost-like spin-2 modes~\cite{Stelle:1977ry}. Consequently, the asymptotic condition~(\ref{infinity}) remains valid throughout the entire weak-field region, not merely at spatial infinity.

\begin{figure}[!htb]
\centering
\includegraphics[width=0.9\linewidth]{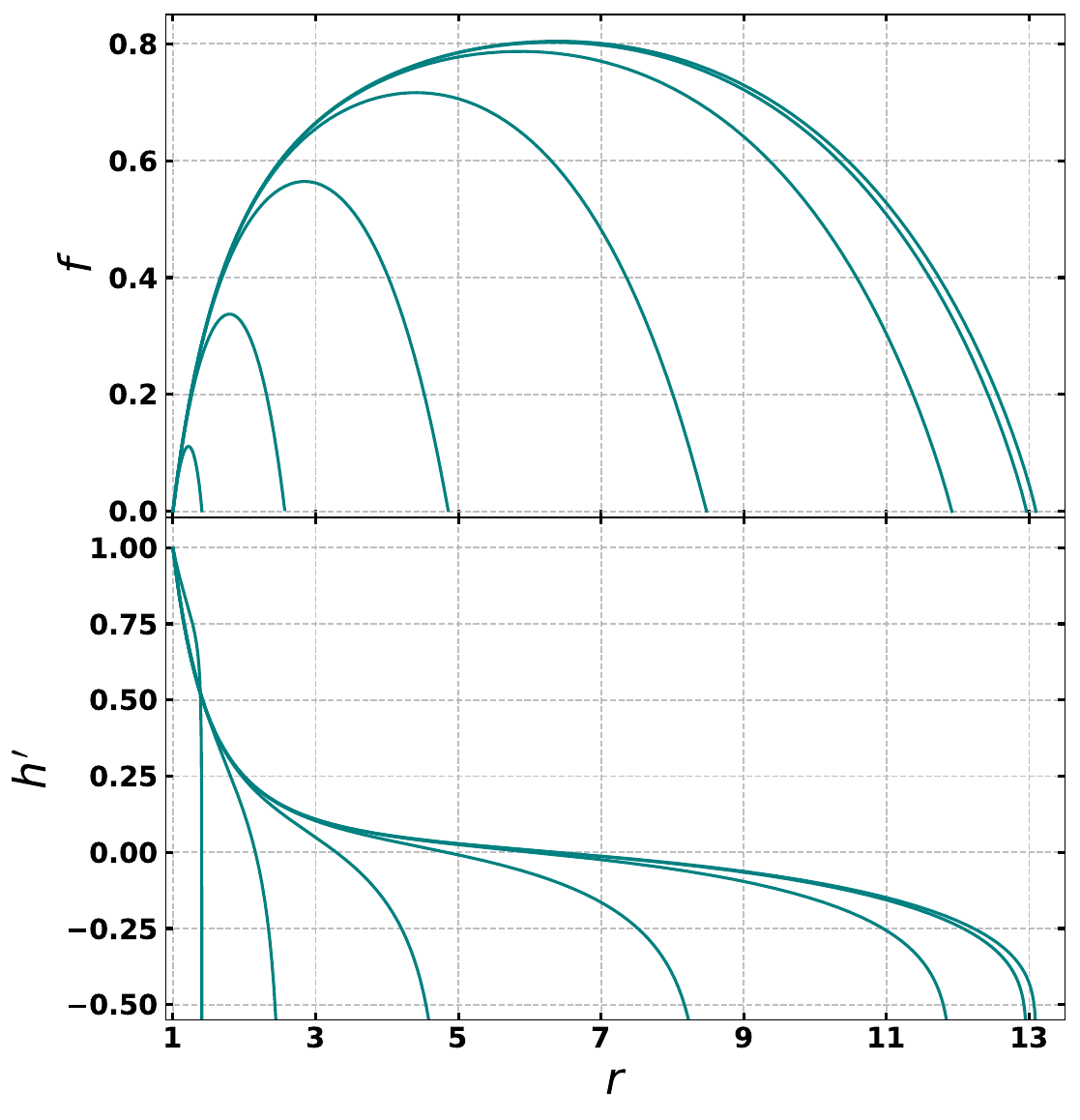}
\caption{\label{fig:mrsly}The solutions of metric functions $f$ and $h'$ for Eqs.~(\ref{eom1})--(\ref{eom2}) with the boundary condition~(\ref{case2}) at event horizon as initial conditions.  The radius of horizon is $r_0 = 1$ and the values of $\lambda$ of the curves from left to right are chosen as $(0.1, 1, 10, 10^2, 10^3, 10^4, 10^5)$.}\label{fplots}
\end{figure}

The near-horizon boundary conditions of Case~I are fully consistent with the asymptotic behavior~(\ref{infinity}), confirming that the Schwarzschild metric represents a global black hole solution of QTG. To test whether Case~II admits a physically acceptable black hole geometry, we numerically integrate the field equations using the corresponding near-horizon data. As shown in Fig.~\ref{fplots}, 
for all values of $\lambda>0$ considered, solutions originating from the Case~II branch fail to approach the asymptotically flat form~(\ref{infinity}). Instead, the metric functions terminate at finite radius, signaling the breakdown of the exterior geometry. We therefore conclude that Case~II does not correspond to a physically viable black hole solution.

The analysis above demonstrates that, although QTG admits a non-unique set of spherically symmetric vacuum solutions, the Schwarzschild spacetime remains the unique static and spherically symmetric black hole solution of the theory, independently of the value of the coupling constant. On the other hand, while vacuum solutions in the strong-field regime are not unique, the weak-field sector of the theory retains uniqueness and coincides exactly with that of GR. As a consequence, QTG naturally avoids constraints from existing weak-field experimental tests on its coupling parameters. These properties establish QTG as an ideal theoretical framework for investigating and comparing the compactness of black holes and compact stars. Since the black hole compactness ${\cal C}$ is fixed at 
\be
{\cal C}_\textup{BH} = \frac{M}{R} =0.5 \,, 
\ee
identical to that in GR, any neutron star configuration attaining ${\cal C}>0.5$ must represent a genuinely novel class of horizonless compact objects. The higher-derivative structure of QTG, together with the absence of ghostlike massive spin-2 modes and the presence of ordinary matter sources, provides a consistent theoretical setting in which such ultra-compact stellar configurations may be supported.

\section{Neutron stars in quasi-topological gravity} \label{NSinQTG}

\subsection{Modified Tolman-Oppenheimer-Volkoff equations}

Having established ${\cal C}=0.5$ as the definitive benchmark for black hole compactness in QTG, we now turn to the equilibrium configurations of neutron stars. Our primary goal is to investigate how QTG modifications allow neutron star compactness to exceed the black hole limit and to characterize the resulting properties of these ultra-compact configurations. In this subsection, we first derive the modified TOV equations within the QTG framework.

Both the interior and exterior spacetimes of a static and spherically symmetric neutron star are described by the metric ansatz given in Eq.~(\ref{ansatz}). To simplify the resulting TOV equations, we introduce the redefinition
\be
h = e^{\gamma} \,,  \label{htogamma}
\ee
which proves convenient for both analytical manipulation and numerical implementation. It is worth noting, however, that when presenting numerical results, $h$ provides a more intuitive visualization of the spacetime properties than $\gamma$, specifically by illustrating its correspondence with $f$. We assume that the stellar interior is composed of a perfect fluid, whose energy-momentum tensor takes the standard form
\be
T^{\mu\nu} = (\rho + p) u^\mu u^\nu + p g^{\mu\nu} \,, \label{tmunu}
\ee
where the density $\rho$ and pressure $p$ depend only on the radial coordinate $r$. These quantities are related by an EOS,
\be
{p} = P ( {\rho} ) \,. \label{eos}
\ee
The four-velocity $u^\mu$ of the fluid is defined as
\be
u^\mu = \frac{dx^\mu}{d\tau} \,, \label{udef}
\ee
with $\tau$ being the proper time, and satisfies the normalization condition
\be
u^\mu u_\mu = -1 \,. \label{ucond}
\ee
Substituting Eqs.~(\ref{ansatz}) and~(\ref{htogamma})--(\ref{eos}) into the gravitational field equations~(\ref{eom}), we find that only the diagonal components $E^i{}_{i}$ are nonvanishing. Their explicit expressions are lengthy and are therefore presented in Appendix~\ref{nonvanishingEOM}. For the general case with $\lambda \neq 0$, by simplifying the non-vanishing components $E^i{}_{i}$, we obtain the modified TOV equations, which form a coupled system of nonlinear ODEs that can be expressed in the general form:
\begin{eqnarray}
\gamma^{\prime\prime\prime}&=& F_1(r, \rho, \gamma^{\prime\prime}, \gamma^{\prime}, f^{\prime\prime}, f^{\prime}, f)\,,\nonumber \\
f^{\prime\prime\prime} &=& F_2(r, \rho, \gamma^{\prime\prime}, \gamma^{\prime}, f^{\prime\prime}, f^{\prime}, f)\,,\nonumber \\
\rho^{\prime}&=& -\frac{(P(\rho) +\rho) \gamma'}{2 P'(\rho)}\,.  \label{tovode}
\end{eqnarray}
Here, the notations $F_1$ and $F_2$ are introduced to simplify the representation of the equations and avoid the display of complex explicit expressions. The third equation in Eq.~(\ref{tovode}) coincides with the continuity equation in GR and directly follows from the conservation of the energy-momentum tensor, $\nabla_\mu T^{\mu\nu} = 0$.

To solve Eq.~(\ref{tovode}), appropriate boundary conditions must be specified. Near the stellar center, regularity requires that the metric functions $\gamma$ and $f$, and matter variable $\rho$ admit power-series expansions of the form
\begin{eqnarray}
\lim_{r\rightarrow 0}\gamma(r) &=& \sum_{n=0}^\infty \gamma_n r^n \,,\nonumber \\
\lim_{r\rightarrow 0}f(r) &=& \sum_{n=0}^\infty f_n r^n \,, \nonumber \\ 
\lim_{r\rightarrow 0}\rho(r) &=& \sum_{n=0}^\infty \rho_n r^n \,. \label{centercond}
\end{eqnarray}
Among the lowest-order coefficients, the central density $\rho_0$ and the parameters $(\gamma_0, \gamma_2, f_2)$ are free, while regularity imposes 
\be
\gamma_1 = f_1 = 0 \,, \quad f_0 = 1\,. 
\ee
For a given EOS and a chosen central density $\rho_0$, the system~(\ref{tovode}) is integrated outward until the stellar surface is reached, which is defined by the vanishing of the pressure,
\begin{equation}
p(R) = 0  \,, \label{surf}
\end{equation}
where $R$ denotes the stellar radius. At the surface, the metric functions are required to be continuous across the interior and exterior regions,
\bea
\gamma_\textup{in}(R) &=& \gamma_\textup{ext}(R) \,, \nn \\
f_\textup{in}(R) &=& f_\textup{ext}(R)  \,,  \label{continue}
\eea
after which the integration proceeds into the exterior vacuum region. Here, the subscripts ``in'' and ``ext'' denote the interior and exterior solutions, respectively. 

In the vacuum exterior, setting $\rho = p = 0$ reduces the field equations to
\begin{eqnarray}
\gamma^{\prime\prime\prime}&=& \tilde{F}_1(r, \gamma^{\prime\prime}, \gamma^{\prime}, f^{\prime\prime}, f^{\prime}, f)\,,\nonumber \\
f^{\prime\prime\prime} &=& \tilde{F}_2(r,  \gamma^{\prime\prime}, \gamma^{\prime}, f^{\prime\prime}, f^{\prime}, f)\,,  \label{vacode}
\end{eqnarray}
where the notations $\tilde{F}_i$ correspond to the source-free limits of $F_i$ defined in Eq.~(\ref{tovode}). At spatial infinity, the metric functions asymptotically approach the Schwarzschild form given in Eq.~(\ref{infinity}).

With the equations~(\ref{tovode}) and~(\ref{vacode}), together with the boundary conditions at the center~(\ref{centercond}), the surface~(\ref{surf}), and spatial infinity~(\ref{infinity}), the problem is well defined once an EOS is specified. We treat the system as a boundary-value problem and employ a numerical shooting method to ensure that the interior solutions match the required exterior vacuum asymptotics.

For numerical convenience, we recast all equations into a dimensionless form by introducing characteristic scales denoted by the subscript ``$\star$''. These scales satisfy
\be
M_\odot \sim r_\star \sim  p_\star^{-\frac12} \sim \rho_\star^{-\frac12} \sim \lambda_\star^{\frac14}   \,, \label{dimension}
\ee
with the solar mass $M_\odot$ chosen as the reference unit. The corresponding dimensional quantities, expressed in centimeter-gram-second (CGS) units, are given by
\bea
r_\star &=& \frac{G M_{\odot}}{c^2} = 1.48 \times 10^{5}\text{cm} \,, \nn \\
\rho_\star &=& \frac{c^6}{G^3 M_{\odot}^2} = 6.18 \times 10^{17}\text{g}\cdot\text{cm}^{-3} \,,  \nn \\ 
p_\star &=& \frac{c^8}{G^3 M_{\odot}^2} = 5.55 \times 10^{38}\text{g}\cdot\text{cm}^{-1}\cdot\text{s}^{-2} \,,  \nn \\
\lambda_\star &=& \frac{G^4 M_{\odot}^4}{c^8} = 4.76 \times 10^{20} \textup{cm}^4 \,. \label{cgs}
\eea

\subsection{Equilibrium configurations}

In this subsection, we investigate the global structure of neutron stars within the framework of QTG. Before presenting the numerical results, it is essential to clarify two closely related points. First, the emergence of neutron star configurations more compact than black holes is fundamentally a higher-derivative strong-field effect induced by the modified gravitational sector, rather than a consequence of subtle nuclear microphysics. Within GR, different EOSs, reflecting different nuclear interactions, can indeed lead to quantitative variations in the stellar mass, radius, and compactness. However, no physically reasonable EOS is able to violate the black hole compactness bound ${\cal C}=0.5$, which is rooted in the geometric structure of GR itself. In QTG, while the black hole compactness remains fixed at ${\cal C}=0.5$, the geometric constraints that limit the compactness of horizonless objects are relaxed, allowing neutron star configurations to exceed the black hole bound. Second, although gravity determines the global spacetime structure of neutron stars, it plays a negligible role in the microscopic nuclear physics encoded in the EOS. The interactions governing dense matter are dominated by the strong nuclear force, in particular short-range repulsive effects. Even at the stellar center, the variation of the spacetime metric across a typical interparticle distance is extremely small, so that local physics is well approximated by special relativity, as if spacetime were flat. For this reason, EOSs are conventionally constructed in flat spacetime and are largely independent of the underlying gravitational theory~\cite{Glendenning:1997wn}. In this context, exploring the impact of gravitational modifications on neutron star configurations~\cite{Olmo:2019flu} while fixing the EOS provides a minimal and controlled setup in which the emergence of ultra-compact, horizonless configurations can be unambiguously attributed to genuine strong-field modifications of gravity, rather than to variations in the modeling of dense matter.

\begin{figure*}[!htbp]
\centering
\includegraphics[width=0.9\textwidth]{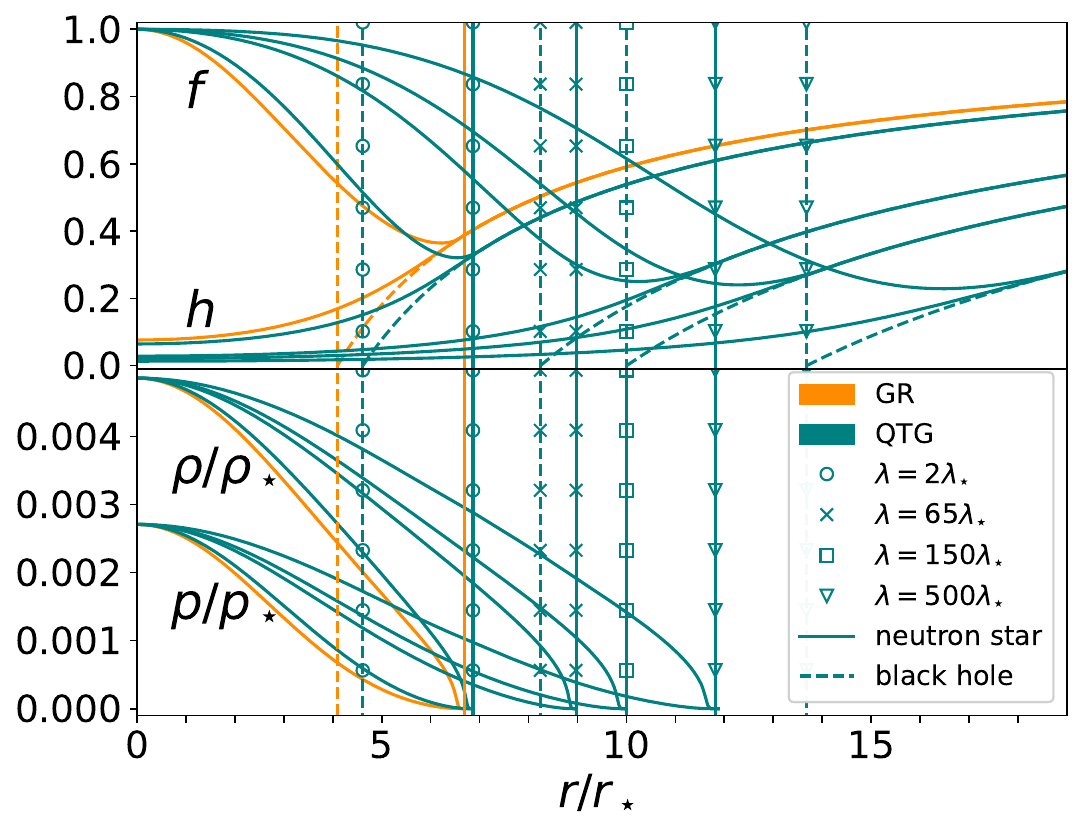}
\caption{Solutions for neutron stars (solid curves) and black holes (dashed curves) in both GR (yellow) and QTG (green). The upper plot presents the numerical solutions for the metric functions $f$ and $h$, showing neutron stars and black holes of the same mass in GR and QTG. The lower panel displays the corresponding radial profiles of density $\rho$ and pressure $p$ for the stellar configurations. We consider QTG coupling constants $\lambda = \{2, 65, 150, 500\} \lambda_\star$. Solid vertical lines indicate the stellar radii $R$, while dashed vertical lines mark the event horizons of the corresponding black holes. All stellar models employ the SLy EOS with a fixed central density of $\rho_0 = 3 \times 10^{15} \, \text{g/cm}^3$ ($4.86 \times 10^{-3} \rho_\star$). In GR, the resulting stellar mass, radius, and compactness are $2.05\,M_{\odot}$, $6.70\,r_\star$, and 0.31, respectively. In the QTG framework, the total mass increases monotonically with the coupling strength, yielding $2.31 M_{\odot}$, $4.12 M_{\odot}$, $5.00 M_{\odot}$, and $6.84 M_{\odot}$ for the aforementioned values of $\lambda$, respectively. The associated stellar radii are $6.86\,r_\star$, $8.97\,r_\star$, $9.98\,r_\star$, and $11.82\,r_\star$, with compactness values 0.34, 0.46, 0.50, and 0.58, respectively.}\label{fig:solution}
\end{figure*}

To isolate the QTG effects, we first examine the spacetime and matter profiles for a representative stellar model constructed using the realistic SLy~\cite{Douchin:2001sv,Haensel:2004nu} EOS with a fixed central density of $\rho_0 = 3 \times 10^{15} \text{g/cm}^3$ across various coupling constants $\lambda$. As shown in the lower panel of Fig.~\ref{fig:solution}, 
the density $\rho$ and pressure $p$ in GR decrease monotonically with the radial coordinate $r$, with the pressure vanishing at the stellar surface. For convenience in evaluating the compactness ${\cal C}$, we present the radial profiles using the dimensionless coordinate $r/r_\star$. Conversion to CGS units can be performed according to Eq.~(\ref{cgs}). When QTG corrections are included, both $\rho$ and $p$ exhibit qualitatively similar radial profiles but decrease more gradually than in the GR case. This suppression of the pressure and density gradients shifts the zero-pressure point to a larger radius, leading to an increased stellar size. At the same time, the enhanced density profile results in a significantly larger total mass. In this regime, hydrostatic equilibrium is maintained through a balance between the modified gravitational attraction and the repulsive pressure gradient. 

The corresponding spacetime structure is illustrated in the upper panel of Fig.~\ref{fig:solution}. In GR, the metric functions $h$ and $f$ are distinct throughout the stellar interior ($h \neq f$) and coincide at the stellar surface, ensuring a smooth matching to the exterior Schwarzschild solution. In contrast, QTG modifications alter the interior geometry such that $h$ and $f$ do not necessarily coincide at the surface itself. Instead, they rapidly converge to equality in the vacuum region just outside the star. As the coupling constant $\lambda$ increases, the radial distance over which this convergence occurs becomes larger. Nevertheless, the convergence always takes place within the strong-field regime, before the spacetime enters the weak-field region. This behavior is guaranteed by the weak-field property of QTG, in which the theory reduces to GR and admits only massless graviton mode.

\begin{figure}[htbp]
\centering
\includegraphics[width=0.9\linewidth]{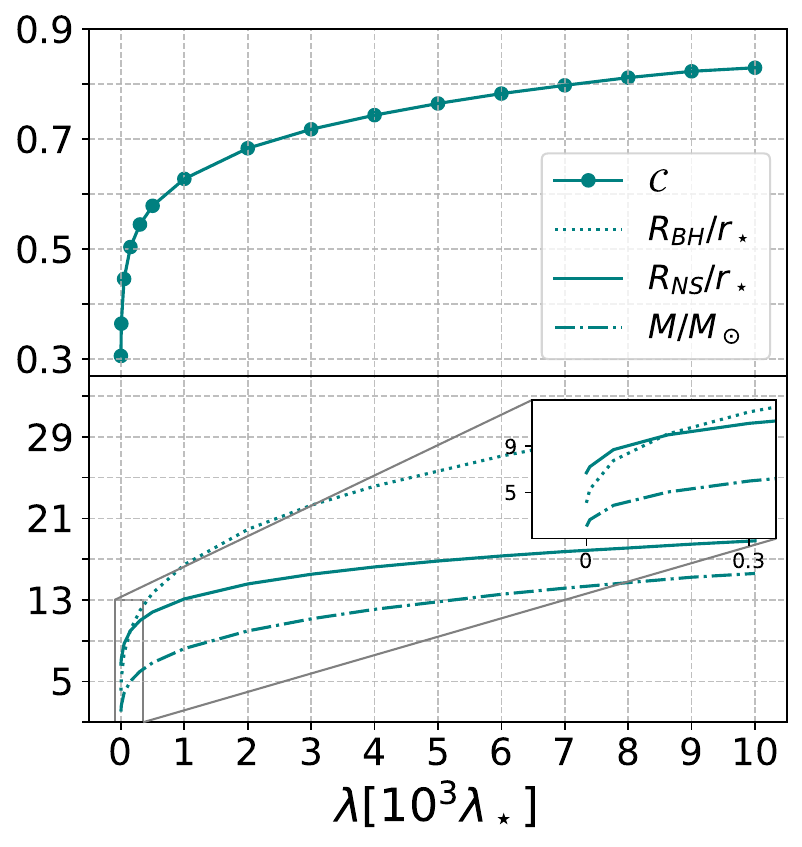}
\caption{The influence of QTG on the compactness $\mathcal{C}$, mass $M$, and radius $R_{\textup{NS}}$ of neutron stars. The left panel shows the ${\cal C}-\lambda$ relation, while the right panel presents the corresponding $M-\lambda$ (dash-dotted line) and $R_{\textup{NS}}-\lambda$ (solid line) relations. The stellar model is constructed using the SLy EOS with a central density $\rho_0 = 3 \times 10^{15}\,\text{g/cm}^3$. For comparison, the Schwarzschild radius $R_{\textup{BH}}$ (dotted line) of a black hole with the same mass $M$ is also shown.}
\label{fig:clambdamr}
\end{figure}

As inferred from Fig.~\ref{fig:solution}, the impact of QTG becomes increasingly pronounced as the coupling constant $\lambda$ increases. To provide a more intuitive and quantitative illustration of how this effect influences the compactness of a given stellar model, we present the dependence of the compactness $\mathcal{C}$ on $\lambda$ over a wider parameter range in the left panel of Fig.~\ref{fig:clambdamr}.
Notably, the $\mathcal{C}-\lambda$ relation is well described by a logarithmic scaling behavior, as confirmed by numerical fitting. The right panel of Fig.~\ref{fig:clambdamr} shows the evolution of mass $M$ and radius $R_{\textup{NS}}$. While both $M$ and $R_{\textup{NS}}$ grow monotonically with $\lambda$, the mass increases more rapidly than the radius, leading to a systematic enhancement of $\mathcal{C}$. This trend can be understood through a simple scaling argument: for a star of uniform density, the radius scales as $R \propto M^{1/3}$, implying $\mathcal{C} = M/R \propto M^{2/3}$. In the absence of a mass upper bound in QTG, the compactness naturally exceeds the black hole limit for sufficiently large masses.

\begin{figure}[ht] 
\centering
\includegraphics[width=0.9\linewidth]{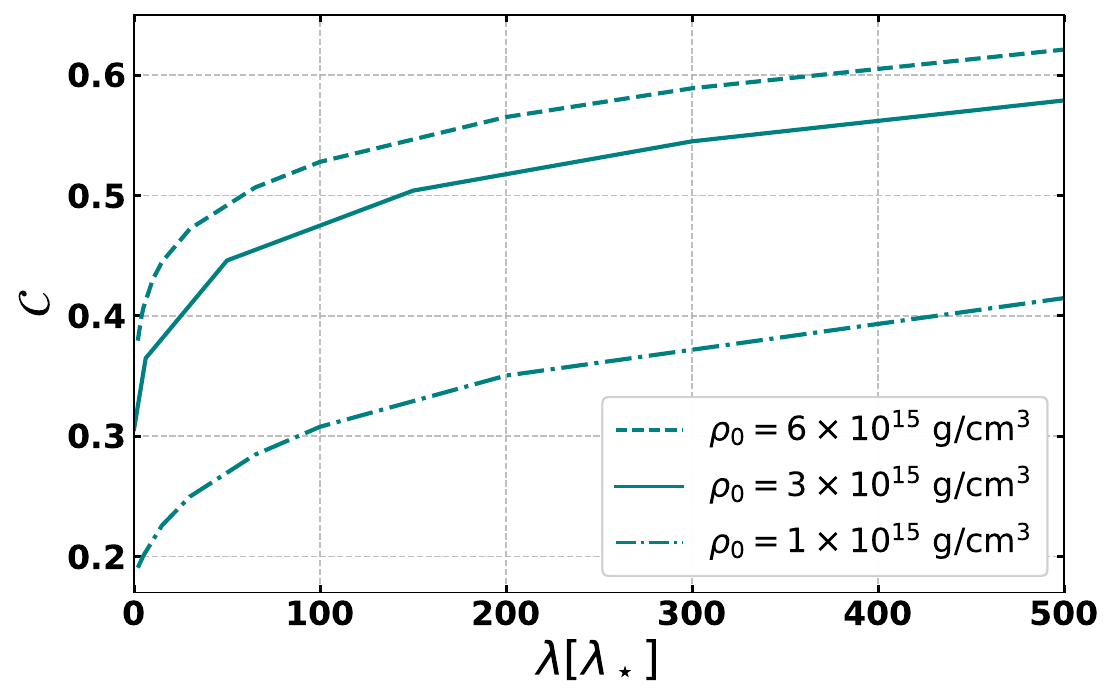}
\caption{The $\mathcal{C}-\lambda$ relation for neutron stars constructed using the SLy EOS, shown for three different central densities: 
$\rho_0 = 6 \times 10^{15}\,\text{g/cm}^3$ (dashed line), 
$\rho_0 = 3 \times 10^{15}\,\text{g/cm}^3$ (solid line), and 
$\rho_0 = 1 \times 10^{15}\,\text{g/cm}^3$ (dash-dotted line).}\label{fig:clambdamr2}
\end{figure}

To demonstrate the universality of this behavior across different central densities, we include two additional stellar models with a lower central density ($\rho_0 = 1 \times 10^{15}\,\text{g/cm}^3$) and a higher central density ($\rho_0 = 6 \times 10^{15}\,\text{g/cm}^3$) for comparison. Figure~\ref{fig:clambdamr2} 
illustrates the corresponding evolution of the compactness $\mathcal{C}$ as a function of the coupling constant $\lambda$. Despite quantitative differences in the absolute values of the compactness, the overall evolutionary trends remain qualitatively similar for all central densities considered. This indicates that the enhancement of compactness induced by QTG  represents a generic feature of the gravitational modifications.

\begin{figure}[!htbp]
\centering
\includegraphics[width=0.9\linewidth]{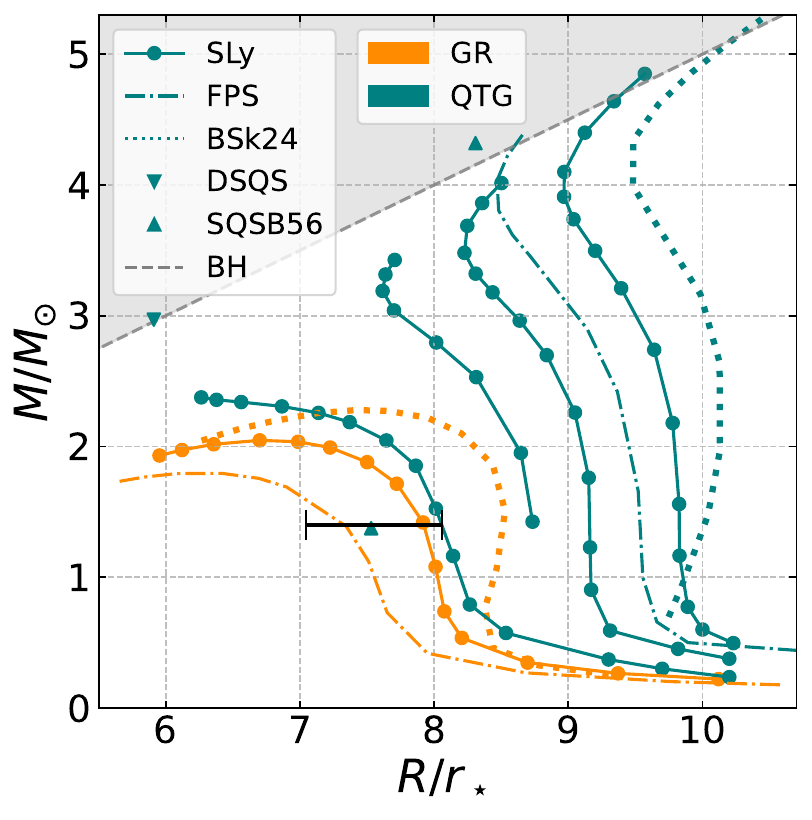}
\caption{The $M-R$ relations for stellar models in GR (yellow) and QTG (green), constructed using different EOSs. For the SLy EOS, the coupling constant from left to right is chosen as $\lambda = 2\lambda_\star$, $15\lambda_\star$, $30\lambda_\star$, and $65\lambda_\star$. For the FPS and BSk24 EOSs, $\lambda$ is fixed at $65\lambda_\star$. For the DSQS and SQSB56 EOSs, the coupling constants are set to $\lambda = 15\lambda_\star$ and $100\lambda_\star$, respectively. The error bar denotes a $1.4\,M_{\odot}$ neutron star with radius $R = 7.45^{+0.61}_{-0.41} r_\star$ ($11.0^{+0.9}_{-0.6}\,\textup{km}$, 90\% credible interval)~\cite{Capano:2019eae}. The black dashed line represents the black hole $M-R$ relation, while the shaded gray region corresponds to stellar configurations with compactness exceeding the black hole limit.}
\label{fig:mr}
\end{figure}

\begin{figure}[ht]
\centering
\includegraphics[width=0.9\linewidth]{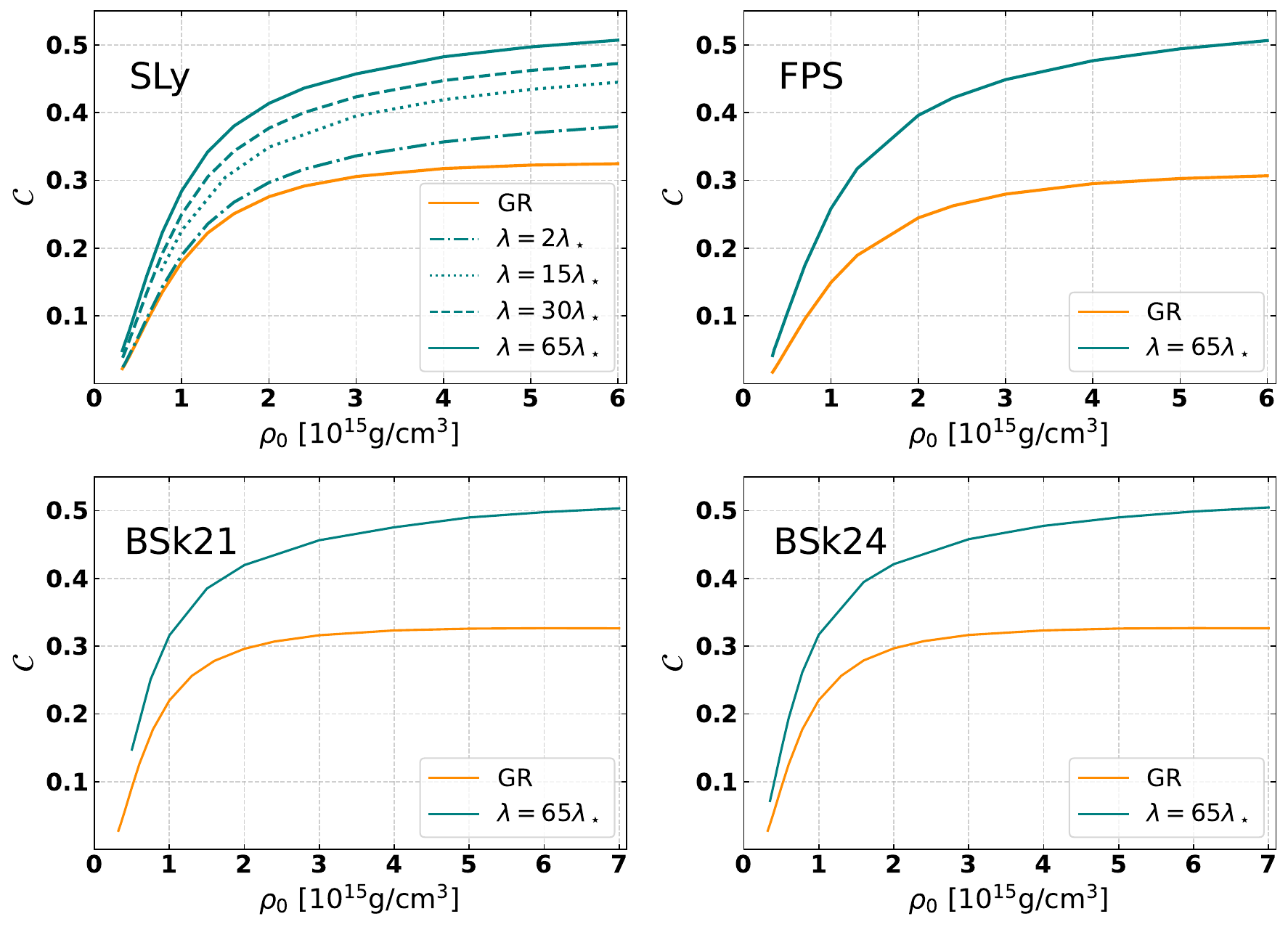}
\caption{The ${\cal C}-\rho_0$ relations for stellar models in GR (yellow) and QTG (green), constructed using different EOSs.}\label{fig:crho0}
\end{figure}

\begin{figure}[ht]
\centering
\includegraphics[width=0.9\linewidth]{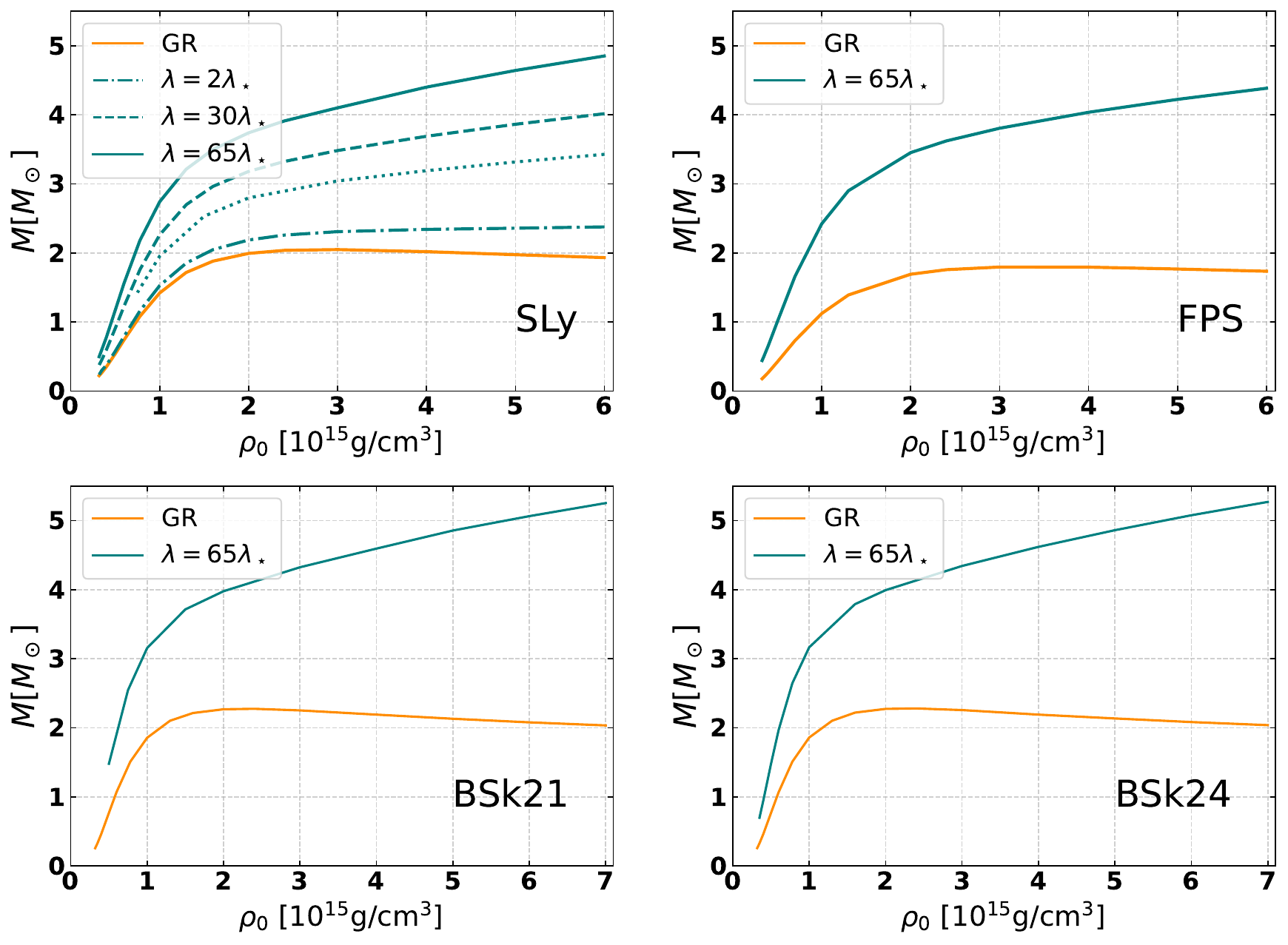}
\caption{The $M-\rho_0$ relations for stellar models in GR (yellow) and QTG (green), constructed using different EOSs.}\label{fig:mrho0}
\end{figure}

To further explore this effect from a global perspective, we plot the mass-radius ($M-R$) relations for neutron star models constructed with the SLy EOS over a representative range of central densities, for several coupling constants $\lambda = 2\lambda_\star,\,15\lambda_\star,\, 30\lambda_\star$, and $65\lambda_\star$, as shown in Fig.~\ref{fig:mr}. 
Several qualitative departures from the GR predictions are immediately evident. In GR, a given EOS admits a maximum stellar mass beyond which no hydrostatic equilibrium configuration exists. In contrast, within the range of central densities considered here, no such upper mass bound is observed in QTG, even for relatively small values of the coupling constant such as $\lambda = 2\lambda_\star$. This already indicates that the higher-curvature corrections can have a nontrivial impact on stellar structure well before entering an extreme coupling regime. At low central densities, QTG neutron stars closely resemble their GR counterparts: the stellar radius decreases as the mass and compactness increase. However, at sufficiently high central densities, a qualitatively new behavior emerges. The stellar radius begins to increase with mass, a trend reminiscent of black hole solutions. Unlike black holes, which strictly obey the linear scaling $R \propto M$ and therefore maintain a fixed compactness $\mathcal{C}=0.5$, QTG neutron stars exhibit a more gradual growth of the radius with mass. As a result, the compactness continues to increase and can exceed the black hole threshold as the central density, and hence the stellar mass, grows further. These features clearly demonstrate that the QTG-induced modifications are intrinsically strong-field effects: they become prominent only in the high-density, high-curvature regime, while being efficiently suppressed in the weak-field region where GR is effectively recovered. The same conclusions can be drawn from complementary perspectives, such as the compactness-central density ($\mathcal{C}-\rho_0$) and mass-central density ($M-\rho_0$) relations, shown in Figs.~\ref{fig:crho0} and~\ref{fig:mrho0}, respectively. 

Although the existence of neutron stars more compact than black holes is not dictated by the nuclear microphysics, examining different EOSs remains instructive. Such a comprehensive investigation not only clarifies the detailed structural behavior of these configurations but also leads to a deeper understanding of the underlying mechanisms governing ultra-compact objects. Isolating these EOS-insensitive properties is essential for understanding neutron stars from a purely gravitational perspective. A well-known example in GR is the I-Love-Q relation~\cite{Yagi:2013bca,Yagi:2013awa,Yagi:2016bkt}, which exhibits a remarkable insensitivity to the EOS and provides a powerful probe of the underlying gravitational theory. In GR, neutron stars are also subject to a maximal mass bound. In QTG, however, this bound is violated, leading to the emergence of configurations that can become more compact than black holes. It is therefore natural to explore whether this behavior persists across different EOSs. Motivated by this consideration, we extend our analysis to additional EOSs, including FPS~\cite{Haensel:2004nu}, BSk21~\cite{Potekhin:2013qqa}, and BSk24~\cite{Pearson:2018tkr}. For these EOSs, we plot the $M-R$, $\mathcal{C}-\rho_0$, and $M-\rho_0$ relations at a representative coupling constant $\lambda = 65\lambda_\star$ in Figs.~\ref{fig:mr}--\ref{fig:mrho0}. Since the results obtained with BSk21 and BSk24 are nearly indistinguishable, only the $M-R$ relation for BSk24 is displayed for clarity. We find that all these EOSs exhibit qualitatively the same behavior as observed for SLy:  the stellar radius increases with mass. 
Crucially, however, the growth rate of the stellar radius remains slower than that of a black hole, so that the compactness ${\cal C}=M/R$ can gradually exceed the black hole bound. This transition occurs at comparable central densities for all EOSs considered. In particular, from the ${\cal C}-\rho_0$ relations shown in Fig.~\ref{fig:crho0}, for the SLy and FPS EOSs, this transition occurs for central densities $\rho_0$ in the range $(5-6)\times10^{15}\textup{g/cm}{}^3$, whereas for the BSk21 and BSk24 EOSs, it occurs slightly higher, in the range $(6-7)\times10^{15}\textup{g/cm}{}^3$. 

If one further considers more exotic forms of dense matter, additional possibilities arise. For example, the $M-R$ relation in Fig.~\ref{fig:mr} shows that for $\lambda = 100\lambda_\star$ and the SQSB56 EOS~\cite{Gondek-Rosinska:2008zmv}, ultra-compact stars with compactness exceeding that of black holes appear at high central densities (such as $\rho_0 = 9 \times 10^{15}\,\text{g/cm}^3$), whereas typical $1.4\,M_\odot$ neutron stars at lower density  ($\rho_0 = 5.4 \times 10^{15}\,\text{g/cm}^3$) remain fully compatible with stringent multimessenger constraints~\cite{Capano:2019eae} within the range of central densities permitted by the EOS. EOSs such as DSQS~\cite{Gondek-Rosinska:2008zmv} allow the transition to ultra-compact configurations at comparatively lower masses  ($\rho_0 = 1.67 \times 10^{16}\,\text{g/cm}^3$), rendering neutron stars more compact than black holes plausible candidates for compact objects inferred in astrophysical observations~\cite{LIGOScientific:2020zkf,LIGOScientific:2024elc}, particularly those populating the mass gap between the most heaviest confirmed neutron stars and the lightest observed black holes.

\begin{figure}
\centering
\includegraphics[width=0.9\linewidth]{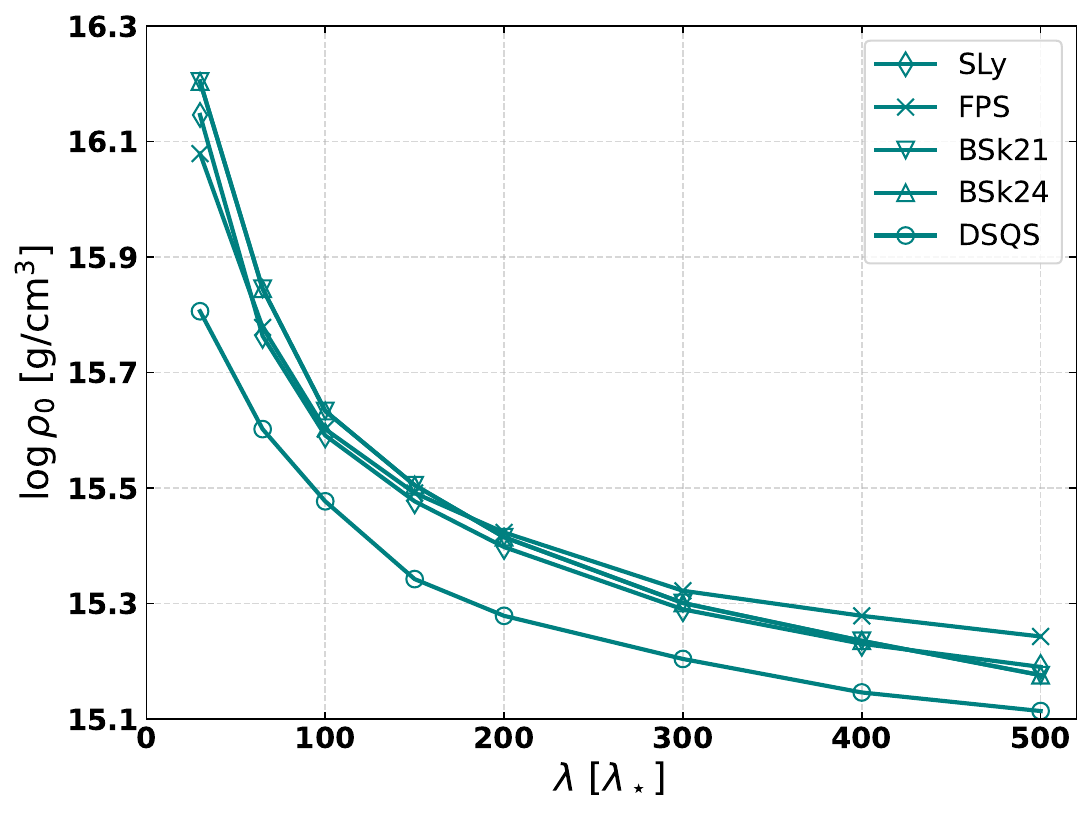}
\caption{Phase diagram in the $(\rho_0, \lambda)$ parameter space.
Different curves correspond to different EOSs (SLy, FPS, BSk21, BSk24, and DSQS).
Each curve represents the critical configuration with compactness $C = 0.5$.
The region above (below) the curve corresponds to configurations with $C > 0.5$ ($C < 0.5$).
}\label{fig:lambdarho0}
\end{figure}

To provide a more intuitive way to explore the dependence of the stellar compactness on both the gravitational coupling and the EOS, we plot a phase diagram in the $(\rho_0, \lambda)$ parameter space in Fig.~\ref{fig:lambdarho0}, 
incorporating five EOSs, including both hadronic models (SLy, FPS, BSk21, BSk24) and a self-bound quark star model (DSQS).
The phase diagram delineates the curves corresponding to the black hole compactness threshold ${\cal C}=0.5$, providing a global view of how ultra-compact configurations emerge as a function of the gravitational coupling and the central density across different EOSs. The region above each curve corresponds to configurations with ${\cal C}>0.5$, while the region below corresponds to ${\cal C}<0.5$.
We find that, although the precise location of the transition depends on the EOS, central density $\rho_0$ and gravitational coupling constant $\lambda$, the overall structure of the phase diagram remains qualitatively similar across different models. In particular, the phase boundaries associated with hadronic EOSs cluster closely together, indicating a weak dependence on the EOS within this class and suggesting an approximately EOS-insensitive behavior. In contrast, the quark star EOS exhibits a noticeable shift, allowing the transition to ${\cal C}>0.5$ to occur at relatively lower central densities for a given $\lambda$. This contrast implies that, for ordinary nuclear matter, the emergence of horizonless configurations with black-hole-level compactness is primarily driven by the QTG modification itself, rather than the details of the EOS. However, when more exotic EOSs are considered, this behavior depends quantitatively on the interplay between the EOS stiffness and the coupling strength.

Moreover, because these ultra-compact stars can be more compact than black holes yet  remain free of both event horizons and spacetime singularities, they may also provide viable alternatives to certain astrophysical sources that are currently interpreted as stellar-mass black holes. This possibility is further encouraged by the fact that event horizons have not yet been directly observed~\cite{Cardoso:2016rao,Cardoso:2016oxy,Cunha:2018gql,Cardoso:2017cqb,Cardoso:2019rvt} and that the presence of singularities is generally regarded as an undesirable feature in any fundamental theory of gravity.

To conclude this section, we assess the physical plausibility of these ultra-compact stellar configurations.  We first emphasize that QTG is a well-defined higher-curvature pure gravity theory constructed from quasi-topological combinations of curvature invariants. The EOSs employed in our analysis are standard, realistic nuclear EOSs. Although the resulting density and pressure profiles differ quantitatively from their GR counterparts, they satisfy the usual physical requirements. Furthermore, we assess the physical plausibility of these ultra-compact configurations from the perspective of energy conditions. 

As emphasized in the Introduction, although certain parameter regions in QTG allow neutron stars to possess radii smaller than their corresponding Schwarzschild radii---without necessarily triggering the formation of trapped surfaces---this behavior does not require a violation of the weak or strong energy conditions. As a representative example, we consider a stellar configuration constructed using the SLy EOS with a central density $\rho_0 = 3 \times 10^{15}\text{g/cm}^3$ and a coupling constant $\lambda = 500\lambda_\star$, whose compactness $\mathcal{C} \approx 0.58$ exceeds the black hole bound ${\cal C}_\textup{BH}$. For this configuration, we explicitly verify that both the weak and strong energy conditions are satisfied throughout the stellar interior as well as in the exterior vacuum region.

\begin{figure}[ht]
\includegraphics[width=0.9\linewidth]{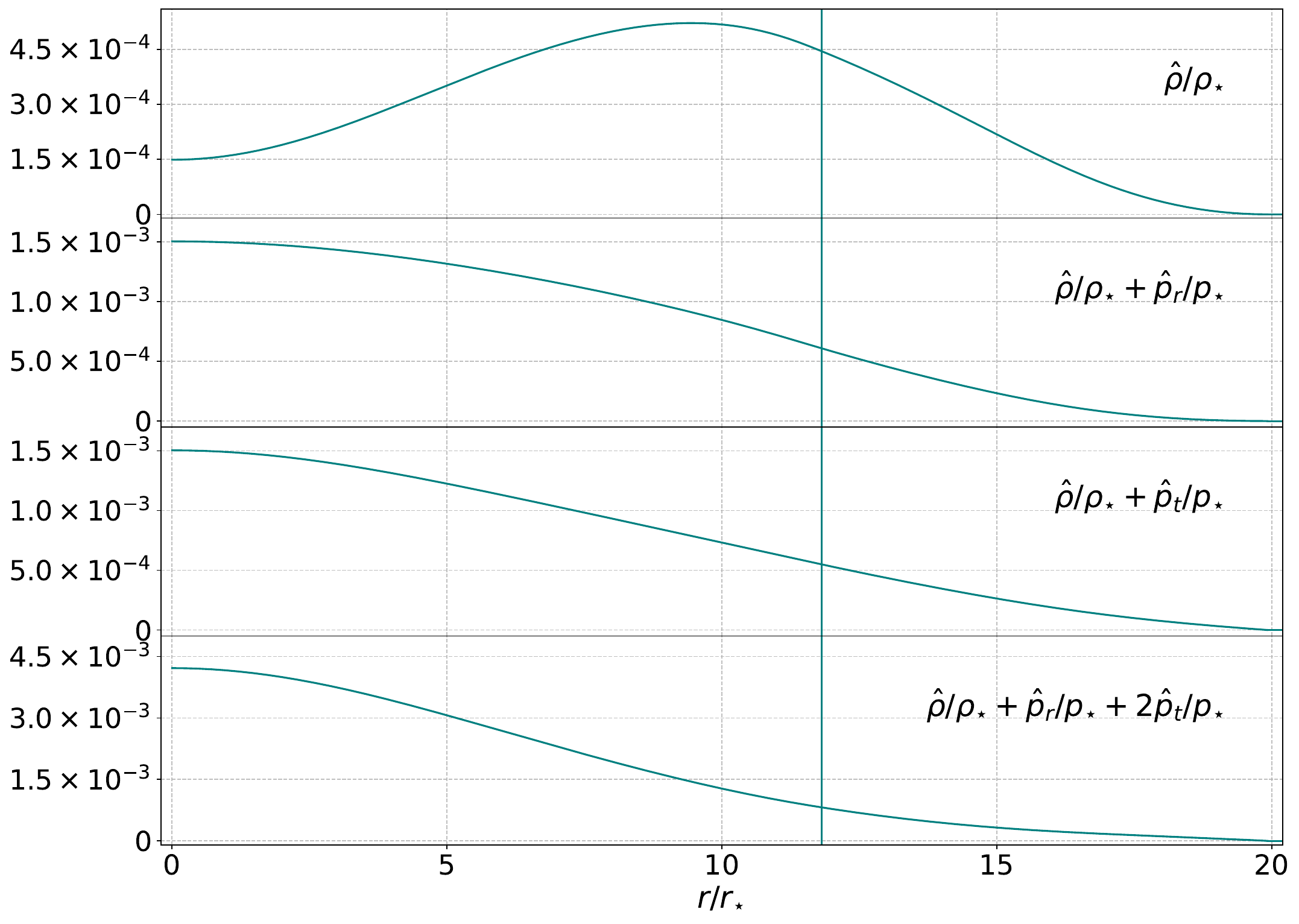}
\caption{\label{fig:energycond} 
Weak and strong energy conditions for the neutron star ($M =6.84 M_{\odot}, R =11.82 r_\star, {\cal C} = M/R\approx 0.58$) more compact than a black hole in QTG ($\lambda =500 \lambda_\star $), where the EOS is SLy and the central density $\rho_0$ is set to $3 \times 10^{15} \textup{g/cm}^3$.
}
\end{figure}

To perform this analysis, we rewrite the gravitational field equations in an effective Einstein-like form,
\be
{G_\mu}^\nu = 8\pi {\hat{T}_\mu}{}^\nu \,, 
\ee
where the effective energy-momentum tensor ${\hat{T}_\mu}{}^\nu$ incorporates the high-order gravitational contributions and is defined as:
\be
{\hat{T}_\mu}{}^\nu = \mathrm{diag}\{-\hat{\rho},\, \hat{p}_r,\, \hat{p}_t,\, \hat{p}_t\} \,.
\ee
The weak and strong energy conditions are then expressed in terms of the effective density $\hat{\rho}$, and the radial and tangential pressures $\hat{p}_r, \hat{p}_t$ as follows:
\begin{subequations} \label{ec}
\bea
&&\hat{\rho} \ge 0 \,,  \label{ec1} \\
&&\hat{\rho}/\rho_\star + \hat{p}_r/p_\star \ge 0 \,,  \label{ec2} \\ 
&&\hat{\rho}/\rho_\star + \hat{p}_t/p_\star \ge 0 \,,  \label{ec3} \\
&&\hat{\rho}/\rho_\star + \hat{p}_r/p_\star + 2\hat{p}_t/p_\star \ge 0 \,.  \label{ec4}
\eea
\end{subequations} 
Equations~(\ref{ec1})--(\ref{ec3}) correspond to the weak energy condition, Eq.~(\ref{ec4}) to the strong energy condition, and Eqs.~(\ref{ec2})--(\ref{ec3}) also constitute the null energy condition. The radial profiles for these conditions are plotted in Fig.~\ref{fig:energycond}.

We note, however, that within this effective curvature fluid description the dominant energy condition is not satisfied, and the effective density $\hat{\rho}$ may exhibit a non-monotonic radial profile. This indicates that the effective curvature fluid does not obey the usual causal condition. While such behaviors may appear concerning at first sight, several clarifications are in order.

First, the procedure of shifting all non-Einstein contributions to the right-hand side of the field equations and interpreting them as an effective energy-momentum tensor is well suited for matter fields minimally coupled to GR, but is generally not appropriate for non-minimally coupled or higher-curvature gravity theories. A well-known example is Starobinsky gravity~\cite{Starobinsky:1980te}, which is widely regarded as a physically viable higher-derivative theory. In its pure curvature formulation, or equivalently in its scalar-tensor representation in the Jordan frame~\cite{Capozziello:2015yza,Liu:2024wvw}, neutron star solutions up to the maximum mass are generally regarded as physically well behaved. Nevertheless, when the same effective curvature fluid approach is employed to examine the energy conditions, one finds that all energy conditions are violated. By contrast, after performing a conformal transformation to the Einstein frame~\cite{Yazadjiev:2014cza}, where the theory is recast as a minimally coupled scalar-tensor gravity, all energy conditions except the strong energy condition are restored. Since Ref.~\cite{Jakubiec:1988ef} demonstrated that any pure gravity theory constructed solely from the Ricci tensor is dynamically equivalent to a scalar-tensor theory, a fully consistent assessment of energy conditions in QTG may likewise require identifying its dynamically equivalent scalar-tensor representation in the Einstein frame, within which the apparent violation of the dominant energy condition may be alleviated or even eliminated.

Second, again using Starobinsky gravity as an illustrative example, one finds that even in the Einstein-frame scalar-tensor formulation, the coupling between the scalar field and matter may lead to an effective density $\hat{\rho}$ that is not monotonic in radius. Therefore, the appearance of a non-monotonic effective density profile in higher-derivative gravity theories does not, by itself, signal any intrinsic pathology of the corresponding stellar solutions.

Third, and more importantly, the emergence of an exotic effective curvature fluid does not by itself imply that ultra-compact stellar configurations can be arbitrarily engineered by postulating an unconventional effective EOS in GR. Even if one were to introduce an ad hoc effective fluid with identical radial profiles of density and pressure, the absence of the specific higher-derivative gravitational dynamics would, in general, preclude the existence of self-consistent equilibrium solutions of this type. In this sense, the ultra-compact stars found in QTG are not artifacts of the effective-fluid description, but genuine solutions arising from the underlying modified gravitational field equations.

Therefore, we argue that assessing the physical viability of stellar configurations in higher-derivative gravities solely from the perspective of energy conditions is not decisive. A more meaningful and robust diagnostic is provided by a systematic analysis of dynamical stability under physically relevant perturbations. Indeed,  Ref.~\cite{Li:2025gna} investigated the stability of neutron stars against radial oscillations in Starobinsky gravity within the Jordan frame and found that, despite violations of all standard energy conditions, the stability properties remain identical to those in GR, with the onset of instability occurring precisely at the maximum-mass configuration. The same conclusion persists after transforming to the Einstein frame, where the effective density may still exhibit non-monotonic radial behavior, yet the stellar configurations remain dynamically stable.

Motivated by these considerations, in the next section we follow the methodology of Ref.~\cite{Li:2025gna} and perform a detailed analysis of radial oscillation stability for neutron stars in QTG. This provides a more definitive and physically meaningful assessment of the viability of the ultra-compact configurations identified in this work.

\section{Radial oscillations of neutron stars} \label{Stability}

In this section, we investigate the stability of neutron stars against radial oscillations. In various astrophysical processes, such as collapse or external disturbances, neutron stars may be subjected to perturbations that excite oscillatory motion of the stellar fluid, giving rise to normal modes. As a first step in perturbation analysis, we focus on the simplest case of radial oscillations. It is worth emphasizing that, in higher-derivative gravity theories, even the analysis of radial oscillations is technically nontrivial. In GR, radial perturbations of a static star do not lead to dynamical excitations in the exterior vacuum spacetime~\cite{Misner:1973prb,Chandrasekhar:1964zza,Kokkotas:2000up}. By contrast, in higher-derivative gravity the exterior geometry generally responds dynamically to such perturbations. In some related studies~\cite{Sotani:2004rq,Sotani:2010re,Staykov:2015cfa}, this difficulty has been circumvented by adopting the Cowling approximation, in which perturbations of the spacetime metric are neglected and only fluid perturbations are retained. However, such an approximation is reliable only when the gravitational modifications are sufficiently small, and it is generally inadequate for probing genuine strong-field effects arising from higher-derivative terms. In Ref.~\cite{Li:2025gna}, a fully self-consistent treatment of radial adiabatic oscillations of neutron stars in higher-derivative gravity was carried out without invoking the Cowling approximation, using Starobinsky gravity as a representative example in both the Jordan and Einstein frames, as well as its Gauss-Bonnet extension~\cite{Liu:2020yqa}. The same approach was also employed in Ref.~\cite{Li:2023vbo} to examine the radial stability of a representative neutron star configuration more compact than a black hole. In the present work, we adopt the same methodology to provide a more detailed analysis of the radial oscillation stability of neutron stars in QTG.

\subsection{Perturbation equations}

To begin, we derive the governing equations for radial perturbations. Under purely radial adiabatic perturbations, the stellar configuration preserves spherical symmetry. Consequently, both the interior and exterior spacetimes can still be described by the metric ansatz given in Eq.~(\ref{ansatz}), with the metric function $h$ replaced by $e^\gamma$ according to Eq.~(\ref{htogamma}). However, the functions $\gamma, f, \rho$, and $p$ are no longer static. Consider a fluid element initially located at radial coordinate $r$ in the equilibrium configuration. Under perturbation, it is displaced to $r + \epsilon \delta r(t, r)$ where $\epsilon$ is a bookkeeping parameter, and  $\delta r$ represents the radial displacement. The perturbed quantities---namely, metric functions ${\gamma}$ and ${f}$, pressure ${p}$,  density ${\rho}$, and four-velocity ${u}^\mu$---are expanded to first order in $\epsilon$ as
\begin{subequations} \label{functionpert}
\bea
{\gamma} &=& \gamma(r) + \epsilon \delta \gamma (t, r) \,, \label{gammapert} \\
{f} &=& f(r) + \epsilon \delta f (t, r) \,,  \label{fpert} \\
{p} &=& p(r) + \epsilon \delta p (t, r) \,,  \label{ppert} \\
{\rho} &=& \rho(r) + \epsilon \delta \rho (t, r) \,, \label{rhopert}  \\
{u}^\mu &=& u^\mu(r) + \epsilon \delta u^\mu (t, r) \,. \label{upert}
\eea
\end{subequations}
Here, $\gamma(r), f(r), p(r), \rho(r)$, and $u^\mu(r)$ denote the equilibrium background quantities. The symbols with ``$\delta$'' represent Eulerian perturbations~\cite{Misner:1973prb}, i.e., variations measured at fixed spatial coordinates. For notational simplicity, we will henceforth use $\gamma, f, p, \rho$, and $u^\mu$ to denote the background quantities without risk of ambiguity. Using Eqs.~(\ref{udef}) and (\ref{upert}), we obtain
\be
\frac{{u}^r}{{u}^t} = \frac{dr/d\tau}{dt/d\tau} = \frac{dr}{dt}= \epsilon \dot{\delta r} \,, \quad \frac{{u}^\theta}{{u}^t} =  \frac{{u}^\varphi}{{u}^t} = 0 \,,\label{uvelocity}
\ee
where the dot denotes the derivative with respect to $t$. Substituting into Eq.~(\ref{ucond}) and expanding to first order in $\epsilon$, we obtain
\bea
u^\mu &=&  (e^{-\frac{\gamma}{2}}, 0, 0, 0) \,, \label{velocity0} \\
\delta u^\mu &=& (-\frac{e^{-\frac{\gamma}{2}}}{2} \delta \gamma, e^{-\frac{\gamma}{2}} \dot{\delta r}, 0, 0)\,. \label{velocity1}
\eea
Using the EOS~(\ref{eos}) and expanding pressure and density as in Eqs.~(\ref{ppert})--(\ref{rhopert}), we have
\bea
p &=& P(\rho) \,, \label{eos0}\\
\delta p &=& \frac{\partial p}{\partial \rho} \delta \rho \label{eos1}\,.
\eea
Substituting the perturbed quantities into the field equations~(\ref{eom}) and expanding in powers of $\epsilon$, the zeroth-order terms reproduce the background equilibrium equations, while the first-order terms yield the linear perturbation equations governing radial oscillations.

We assume harmonic time dependence for the radial displacement $\delta r$ and all perturbed variables,
\begin{subequations} \label{timedependent}
\bea
\delta r(t, r) &=& \xi (r) e^{i \omega t} \,, \label{rtimedependent}\\
\delta \gamma(t, r) &=& \delta \gamma (r) e^{i \omega t} \,, \label{lambdatimedependent}\\
\delta f(t, r) &=& \delta f (r) e^{i \omega t} \,, \label{ftimedependent}\\
\delta p(t, r) &=& \delta p (r) e^{i \omega t} \,, \label{ptimedependent}\\
\delta \rho(t, r) &=& \delta \rho (r) e^{i \omega t} \,, \label{rhotimedependent}
\eea
\end{subequations}
where $\omega$ is the oscillation frequency.
The ${\cal O}(\epsilon)$ components of the field equations lead to a coupled system of linear ODEs. The explicit expressions of the nonvanishing components are presented in Appendix~\ref{pertNSinQTG}. These equations can be reorganized into the schematic form
\begin{eqnarray}
\xi^{\prime\prime}&=& A_{11} \xi^{\prime} + A_{12} \xi  + A_{13} \delta \gamma^{\prime\prime} + A_{14} \delta \gamma^{\prime}  + A_{15} \delta f^{\prime} + A_{16} \delta f  \,, \nonumber  \\
\delta \gamma^{\prime\prime\prime}&=& A_{21} \xi^{\prime} + A_{22} \xi  + A_{23} \delta \gamma^{\prime\prime} + A_{24} \delta \gamma^{\prime} + A_{25} \delta f^{\prime} + A_{26} \delta f  \,,\nonumber \\
\delta f^{\prime\prime} &=& A_{31} \xi^{\prime} + A_{32} \xi  + A_{33} \delta \gamma^{\prime\prime} + A_{34} \delta \gamma^{\prime}  + A_{35} \delta f^{\prime} + A_{36} \delta f \,, \label{radialpert}
\end{eqnarray}
with $\delta p =  P'(\rho) \delta \rho$, and 
\begin{equation}
\delta \rho = A_{41} \xi^{\prime} + A_{42} \xi  + A_{43} \delta \gamma^{\prime\prime} + A_{44} \delta \gamma^{\prime} + A_{45} \delta f^{\prime} + A_{46} \delta f \,,  
\end{equation}
where a prime denotes the derivative with respect to $r$. The coefficients $A_{ij}$ depend on the background functions $(\gamma, \gamma^{\prime},  \gamma^{\prime\prime}, f, f^{\prime}, f^{\prime\prime}, \rho)$, the radial coordinate $r$, and the parameters $(\lambda, \omega)$. Their explicit expressions are lengthy and non-instructive and are therefore not displayed. 

Physically acceptable solutions must satisfy boundary conditions at the center, the stellar surface, and spatial infinity. Near the center ($r \to 0$), we perform a power series expansion for the perturbation variables $\{\delta \gamma, \delta f, \xi\}$ as
\begin{eqnarray}
\lim_{r\rightarrow 0}\delta \gamma(r) &=& \sum_{n=0}^\infty \delta \gamma_n r^n \,,\nonumber \\
\lim_{r\rightarrow 0}\delta f(r) &=& \sum_{n=0}^\infty \delta f_n r^n \,, \nonumber \\ 
\lim_{r\rightarrow 0}\xi(r) &=& \sum_{n=0}^\infty \xi_n r^n \,, \label{centercond2}
\end{eqnarray}
Regularity at the origin requires 
\be
\delta \gamma_1= \delta f_0= \delta f_1= \xi_0=0 \,,
\ee
while $(\delta \gamma_0, \delta \gamma_2, \delta f_2, \xi_1)$ remain free parameters.  At the stellar surface ($r = R$), the Lagrangian perturbation of the pressure $\Delta p$ must vanish, which is expressed as:
\begin{equation}
\Delta p (R) =  0 \,. \label{surf2}
\end{equation}
Continuity of the metric perturbations requires
\bea
\delta\gamma_\textup{in}(R) &=& \delta\gamma_\textup{ext}(R) \,, \nn \\
\delta f_\textup{in}(R) &=& \delta f_\textup{ext}(R) \,.   \label{continue2}
\eea
Unlike in GR, the exterior spacetime in QTG responds dynamically to radial perturbations. Therefore, the perturbations must be integrated into the vacuum region ($p = \rho = 0$). Asymptotic flatness requires
\begin{equation}
\lim_{r\rightarrow\infty} \delta \gamma(r) = \lim_{r\rightarrow\infty} \delta f(r) = 0 \,. \label{infinity2}
\end{equation}
In practice, owing to the weak-field behavior of QTG, the metric perturbations decay rapidly and approach zero well before entering the asymptotic weak-field regime.

Having specified the perturbation equations and boundary conditions, we close the system by adopting an EOS. 
We treat the problem as a boundary-value eigenvalue problem and employ a numerical shooting method to ensure that the interior solutions match the required exterior vacuum asymptotics. Following the numerical strategy described in detail in Ref.~\cite{Li:2025gna}, we fix $\xi_1 = 1$, which merely sets the overall normalization (amplitude) of the radial displacement and does not affect the eigenvalue $\omega^2$. 
The parameter $\delta\gamma_0$ likewise does not influence the determination of $\omega^2$, since $\delta\gamma$ itself does not explicitly enter the perturbation equations~(\ref{radialpert}).

For numerical convenience, we recast all equations into dimensionless form by further introducing the characteristic frequency scale $\omega_\star$, such that $M_\odot \sim \omega_\star^{-1}$, which is defined as
\begin{equation}
\omega_\star = \frac{c^3}{G M_{\odot}} = 2.03 \times 10^{5}\,\text{s}^{-1} \,.
\end{equation}

\subsection{Stability analysis}

\begin{figure}[ht]
\centering
\includegraphics[width=0.9\linewidth]{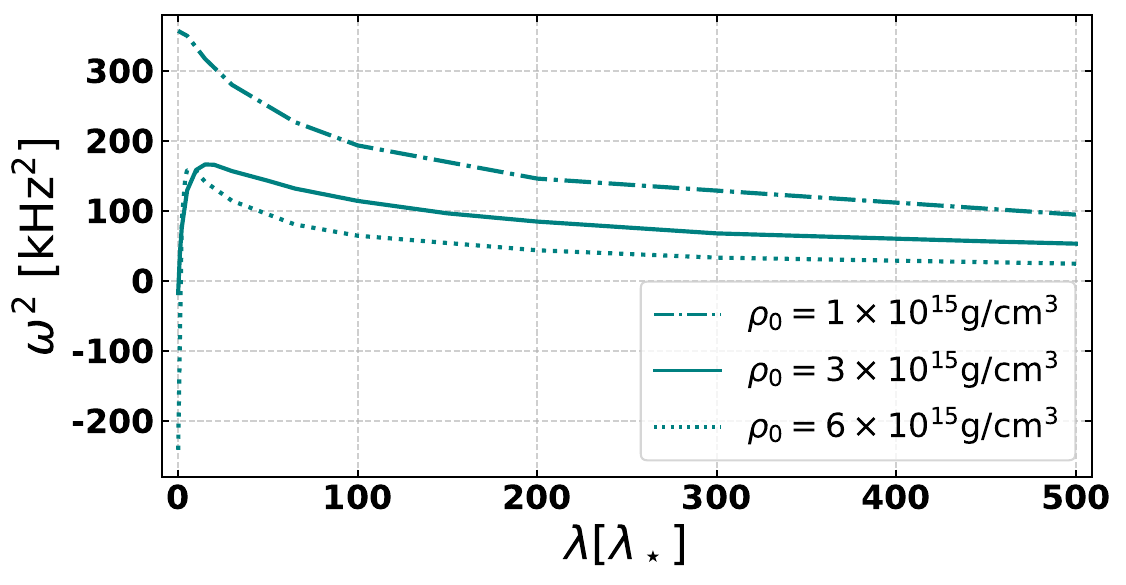}
\caption{The $\omega^2-\lambda$ relation for neutron stars constructed using the SLy EOS, shown for three different central densities: 
$\rho_0 = 6 \times 10^{15}\,\text{g/cm}^3$ (dotted line), 
$\rho_0 = 3 \times 10^{15}\,\text{g/cm}^3$ (solid line), and 
$\rho_0 = 1 \times 10^{15}\,\text{g/cm}^3$ (dash-dotted line).}\label{omegalambda}
\end{figure}

Reference~\cite{Li:2023vbo} examined the radial stability of a representative stellar configuration constructed with the SLy EOS at a central density $\rho_0 = 3 \times 10^{15}\,\text{g/cm}^3$ in both GR and QTG with coupling constant $\lambda = 500\,\lambda_\star$. It was shown that sufficiently strong QTG corrections not only increase the compactness of this configuration beyond the black-hole limit, but can also convert it from radially unstable to stable. We now investigate in more detail how the radial stability of this same stellar configuration evolves as the strength of the QTG corrections varies. In Fig.~\ref{omegalambda}, 
we plot the squared frequency $\omega^2$ of the fundamental mode as a function of the coupling constant $\lambda$. 
As shown in Fig.~\ref{omegalambda}, the stellar model with $\rho_0 = 3 \times 10^{15}\,\text{g/cm}^3$ is radially unstable in GR, as indicated by $\omega^2 < 0$. Once QTG corrections are introduced, however, $\omega^2$ undergoes a sharp transition from negative to positive values, signaling a rapid crossover from an unstable regime to a stable one. Remarkably, this stabilization already occurs for very small coupling strengths, for instance at $\lambda = 2\lambda_\star$. As $\lambda$ increases further, the squared frequency $\omega^2$ reaches a maximum at a finite value of the coupling constant and then decreases gradually. Importantly, $\omega^2$ remains positive throughout this regime, indicating that the stellar configurations remain dynamically stable even in the large-$\lambda$ limit.

For comparison, we also consider stellar models with a higher central density, $\rho_0 = 6 \times 10^{15}\,\text{g/cm}^3$, and a lower central density, $\rho_0 = 1 \times 10^{15}\,\text{g/cm}^3$. Their corresponding $\omega^2-\lambda$ relations are likewise shown in Fig.~\ref{omegalambda}. For the higher-central-density model, the squared frequency in GR is significantly more negative, indicating a stronger instability. Nevertheless, $\omega^2$ crosses zero and becomes positive at even smaller values of $\lambda$ than in the $\rho_0 = 3 \times 10^{15}\,\text{g/cm}^3$ case. After reaching its maximum, $\omega^2$ decreases more rapidly and settles at a lower asymptotic value. In contrast, the low-central-density model is already radially stable in GR. Consequently, no initial growth of $\omega^2$ is observed when QTG corrections are introduced. Instead, its squared frequency exhibits a monotonic decrease with increasing $\lambda$, qualitatively resembling the large-$\lambda$ behavior of the other two models.

\begin{figure}[ht]
\centering
\includegraphics[width=0.9\linewidth]{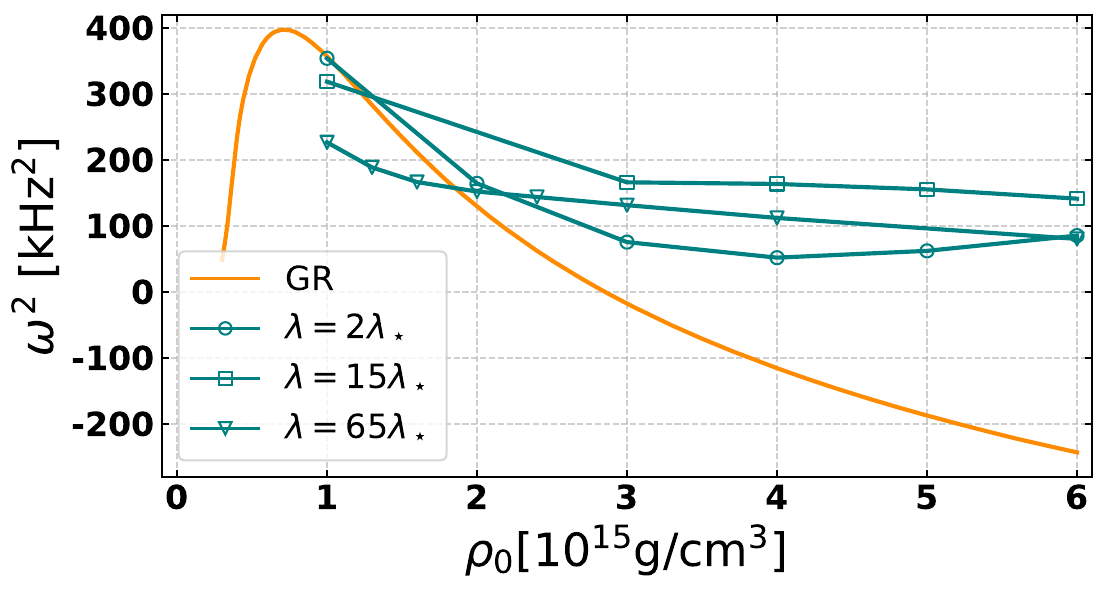}
\caption{The $\omega^2-\rho_0$ relations for stellar models in GR (yellow) and QTG (green), constructed using SLy EOS. The coupling constant is chosen as $\lambda = 2\lambda_\star$, $15\lambda_\star$, and $65\lambda_\star$.}\label{omegarho}
\end{figure}

To further clarify the global impact of QTG corrections, we also plot the $\omega^2-\rho_0$ relations for neutron stars constructed with the SLy EOS at several representative coupling constants, $\lambda = 2\lambda_\star$, $15\lambda_\star$, and $65\lambda_\star$, as shown in Fig.~\ref{omegarho}.
For small QTG corrections, such as $\lambda = 2 \lambda_\star$, the squared frequency $\omega^2$ is nearly insensitive to QTG effects in the low-central-density regime. This behavior is consistent with the trends observed in Fig.~\ref{omegalambda}, and reflects the suppression of higher-derivative corrections in weak-field configurations. As the coupling constant $\lambda$ increases, QTG effects become progressively more significant. The range of central densities over which $\omega^2$ is substantially modified broadens. A particularly noteworthy feature is that the impact of QTG on the squared frequency of high-central-density stellar models is opposite to that found in Starobinsky gravity. In Starobinsky gravity, higher-curvature corrections tend to further destabilize high-central-density configurations, driving $\omega^2$ toward more negative values~\cite{Li:2025gna}. In contrast, QTG corrections can render such configurations dynamically stable, even when they are unstable in GR. Moreover, combining with the $M-\rho_0$ relation shown in Fig.~\ref{fig:mrho0}, we find that the standard GR radial stability criterion, $dM/d\rho_0 > 0$, remains valid in QTG. However, even for very small QTG corrections, e.g., $\lambda = 2\lambda_\star$, the mass curve exhibits no maximum within the typical range of central densities. Consequently, the unstable branch characterized by $dM/d\rho_0 < 0$ disappears.

Finally, we note that all frequencies presented above correspond to the fundamental radial oscillation mode of the stellar models. We now briefly discuss the behavior of overtone modes to conclude this section. In GR, radial oscillations generate perturbations of both the fluid and the spacetime geometry, but these perturbations are confined to the stellar interior. The exterior vacuum spacetime remains unaffected. Inside the star, the perturbation equation reduces to a second-order ODE for $\xi$, which can be cast into the standard Sturm-Liouville form~\cite{Misner:1973prb,Chandrasekhar:1964zza,Kokkotas:2000up}. As a result, the system admits an infinite discrete spectrum of eigenvalues, corresponding to an infinite set of squared frequencies $\omega_n^2$. The lowest eigenvalue represents the fundamental mode ($n=0$), while the higher ones correspond to overtone modes ($n=1, 2, \dots$). In higher-curvature gravity theories, the situation is qualitatively different because the exterior vacuum spacetime generally responds dynamically to perturbations. The perturbation equations must therefore be solved in both the interior and exterior regions. In generic higher-curvature gravities, the vacuum response extends over the entire spacetime, including both strong-field and weak-field regions. In the weak-field regime, the presence of massive scalar modes or massive spin-2 modes typically leads to Yukawa-type behaviors. Only the fundamental mode is usually associated with a decaying solution of the form $e^{-\mu r}$, whereas overtone modes tend to contain growing components of the form $e^{\mu r}$, which violate the boundary condition at spatial infinity. Consequently, overtone modes are generally absent in such theories. This feature has been observed in Starobinsky gravity and in its Gauss-Bonnet extension, where a massive scalar mode is present~\cite{Li:2025gna}.

\begin{figure}[!htbp]
\centering
\includegraphics[width=0.9\linewidth]{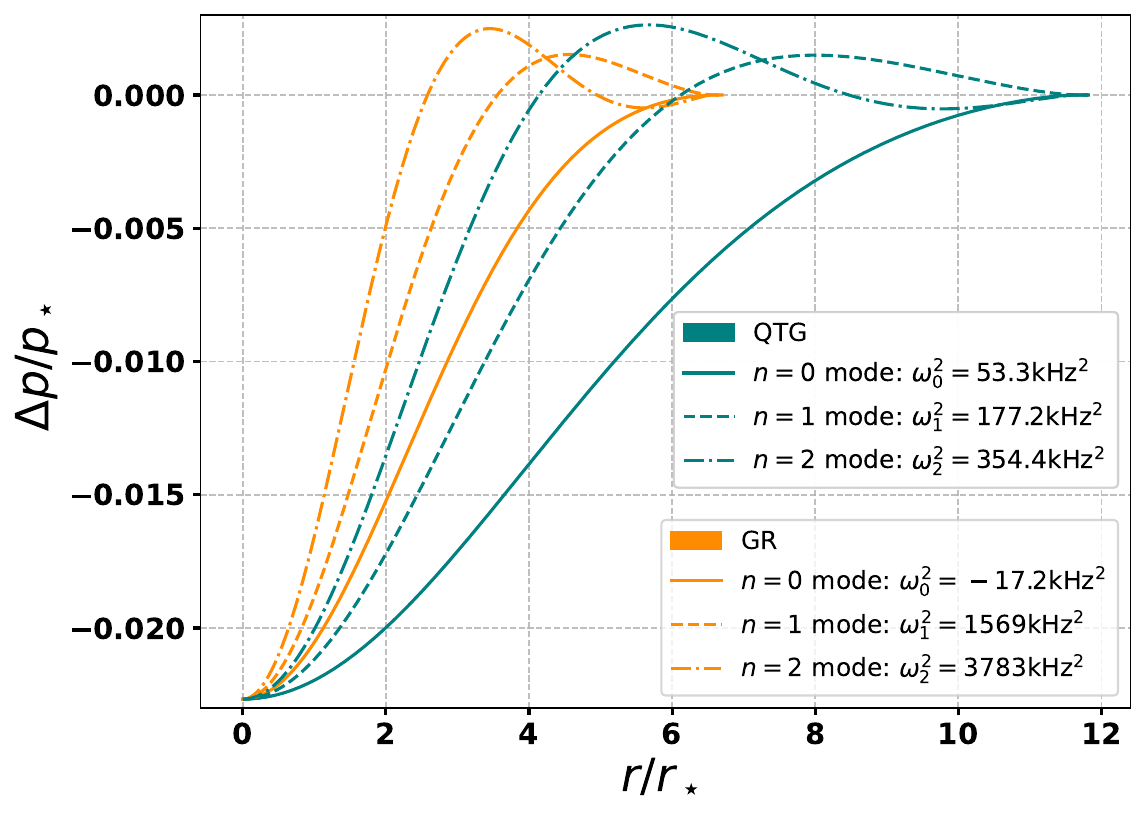}
\caption{Numerical solutions of the Lagrangian perturbations of pressure $\Delta p$ for a stellar model constructed using the SLy EOS with a central density of $\rho_0 = 3 \times 10^{15} \, \text{g/cm}^3$ in GR (yellow) and in QTG (green) with a coupling constant $\lambda = 500 \lambda_\star$. Solid, dashed, and dash-dotted curves correspond to the fundamental mode ($n=0$), the first overtone ($n=1$), and the second overtone ($n=2$), respectively.}\label{overtone}
\end{figure}

As discussed earlier, however, QTG does not introduce additional massive modes in the weak-field regime and reduces smoothly to GR outside the strong-field region. Therefore, once the perturbations propagate into the weak-field domain, both the fundamental mode and the overtone modes decay toward zero in a manner similar to GR. The radial oscillation equations thus reduce to a well-defined Sturm-Liouville-like eigenvalue problem, naturally admitting a discrete spectrum of overtone modes. To verify this argument, we again consider a representative stellar model constructed using the SLy EOS with a central density $\rho_0 = 3 \times 10^{15}\,\text{g/cm}^3$ in QTG, with the coupling constant fixed at $\lambda = 500 \lambda_\star$. For this configuration, we compute the oscillation frequencies for the modes $n=0, 1,$ and $2$, where $n$ denotes the number of nodes inside the star. Here, $n=0$ corresponds to the fundamental mode, while $n=1$ and $n=2$ represent the first and second overtones, respectively. The corresponding Lagrangian perturbations of pressure $\Delta p$ are plotted in Fig.~\ref{overtone}.
One can see that although the QTG correction drives the squared frequency of the fundamental mode from negative to positive values, signaling stabilization, the squared frequencies of the overtone modes remain noticeably smaller than their GR counterparts.

\section{Conclusions} \label{Conclusion}

In this work, we have carried out a detailed investigation of the equilibrium configurations and radial oscillation stability of neutron stars more compact than black holes, as proposed in Ref.~\cite{Li:2023vbo}, within the framework of QTG. Our main findings are summarized as follows.

First, regarding  the equilibrium structure,  we considered several representative EOSs and different values of gravitational coupling constant and found that, in the high-central-density regime,  the resulting  mass-radius relations exhibit a universal behavior reminiscent of black holes, namely that the radius increases with mass. Unlike black holes, however, the stellar radius grows more slowly with increasing mass, so that the compactness ${\cal C}=M/R$ can gradually exceed the black hole value. These results indicate that the emergence of ultra-compact neutron star configurations in QTG is a universal qualitative feature, while its quantitative realization depends sensitively on the magnitude of the higher-derivative gravitational coupling, the choice of EOS, and the central density.

Second, regarding radial stability, our analysis shows that QTG corrections become increasingly significant at higher central densities. In particular, stellar configurations that are radially unstable in GR can be stabilized once QTG effects are taken into account, regardless of whether their compactness exceeds the black-hole bound. This stabilization occurs even for relatively small values of the coupling constant and persists over a broad region of parameter space.

Third, in comparison with Starobinsky gravity, where additional massive modes in the weak-field regime modify the perturbation spectrum and typically remove overtone modes, QTG exhibits a qualitatively different behavior. Because QTG reduces to GR in the weak-field limit without introducing extra massive propagating degrees of freedom, both the fundamental and overtone radial oscillation modes are preserved. The perturbation problem therefore retains a well-defined discrete spectrum analogous to that in GR, while allowing for substantially different behavior in the strong-field regime.

Several remarks are in order regarding equilibrium configurations and stability of these ultra-compact stellar models. First, it is worth exploring whether similarly ultra-compact, horizonless objects can arise in other alternative theories of gravity. Although QTG provides a concrete and technically tractable framework in which neutron stars more compact than black holes can be constructed and systematically analyzed, it needs not constitute a unique realization of this strong-field phenomenon. The quasi-topological Ricci cubic gravity considered in the present work represents the simplest member within a broader class of QTG theories~\cite{Li:2017ncu,Hennigar:2017ego,Chen:2022fdi}. It remains to be investigated whether models involving explicit Riemann tensor contributions, as well as higher-order extensions such as quintic Ricci terms, may yield similar or even more pronounced ultra-compact configurations. Moreover, since Ref.~\cite{Jakubiec:1988ef} demonstrated that any pure gravity theory constructed solely from the Ricci tensor is dynamically equivalent to a scalar-tensor theory, this equivalence suggests an alternative avenue for exploring ultra-compact objects within nonminimally coupled gravity frameworks. 

Second, it is essential to examine the stability of these ultra-compact neutron stars under more general and complex perturbations. Since QTG corrections not only increase the compactness of stellar configurations but can also render radially unstable GR solutions stable, it is natural to explore whether similar stabilizing effects may arise in other instability channels identified within GR~\cite{Cardoso:2019rvt,Friedman:1978ygc,Comins1,Yoshida1,Kokkotas:2002sf,Friedman:1978wla,Cardoso:2007az,Brito:2015oca,Maggio:2017ivp,Maggio:2018ivz,Keir:2014oka,Cardoso:2014sna,Cunha:2022gde}. Such analyses are, however, technically more involved and may reveal qualitatively new dynamical features and sharpen the observational viability of these horizonless, ultra-compact remnants.

Finally, we briefly comment on the potential observational implications of the ultracompact configurations obtained in this work. Our results show that QTG admits horizonless stellar configurations with compactness exceeding the black hole limit, while remaining radially stable. This raises the possibility that such objects could serve as candidates for compact objects residing in the mass gap and mimic stellar-mass black holes in certain observational channels, while still exhibiting characteristic signatures.
On the one hand, in the post-merger and ringdown stages of compact binary coalescence events, horizonless ultracompact objects may give rise to gravitational-wave echoes due to the presence of an effective surface or partially reflective structure~\cite{Cardoso:2017cqb,Cardoso:2019rvt,Cardoso:2016rao,Cardoso:2016oxy}. Moreover, the echo frequency is expected to decrease as the compactness increases~\cite{Li:2023vbo}. A more detailed characterization of such echo signals, including their frequency spectrum and damping properties, would require a systematic analysis of the dynamical response of these stars under perturbations.
On the other hand, during the inspiral phase of a compact binary, additional observables may provide complementary probes. In particular, tidal deformability~\cite{Cardoso:2017cfl}, tidal heating~\cite{Maselli:2017cmm}, and the multipolar structure~\cite{Krishnendu:2017shb} can carry imprints of the internal structure of the compact object. In contrast to black holes in GR, which have vanishing tidal Love numbers and exhibit tidal heating associated with the presence of an event horizon, horizonless ultracompact objects can possess nonvanishing tidal deformability and distinct multipolar structures while lacking horizon-induced dissipation~\cite{Cardoso:2017cqb,Cardoso:2019rvt,Berti:2015itd,Barack:2018yly,LISA:2022kgy}.
Taken together, these observables may provide a consistent framework---potentially in the form of universal relations---to distinguish ultracompact stars in QTG from classical black holes or other exotic compact objects in GR or alternative theories of gravity. With the ongoing improvement in gravitational-wave and electromagnetic observations, as well as the advent of next-generation detectors, these signatures may become accessible in the near future. This would open a new window to probe the nature of compact objects in the strong-field regime and to test possible deviations from general relativity.

\appendix 

\section{Non-vanishing components of equations of motion} \label{nonvanishingEOM}

For completeness, we provide the explicit expressions for the non-vanishing components of the field equations~(\ref{eom}) below:
\begin{widetext}
\begin{subequations} \label{TOV}
\bea
E^t{}_t &\equiv& \frac{6 f \lambda f''' \left(f'-f \gamma '\right)}{r^2} -\frac{3 f \lambda  \gamma '' \left(4 f^2 \left(r^2 \gamma '{}^2-1\right)+f \left(4-5 r^2 \gamma ' f'\right)+r^2 f'{}^2\right)}{r^4}  \nn \\
%%%%%%%%%%
&\quad&  +\frac{6 f \lambda f''^2}{r^2}+ \lambda f'' \left[\frac{3   \left(2 f^2 \left(r^2 \gamma '{}^2-2\right)+f \left(4-5 r^2 \gamma ' f'\right)+r^2 f'{}^2\right)}{r^4}-\frac{6 f^2 \gamma ''}{r^2}\right]  \nn\\
%%%%%%%%%%
&\quad& -3 \lambda  \bigg[f^3 \gamma ' \left(r^3 \gamma '{}^3-2 r \gamma '+16\right)+f f' \left(4 \left(3 r \gamma '+4\right)-7 r f' \left(r^2 \gamma '{}^2-2\right)\right) \nn\\
%%%%%%%%%%
&\quad& +r f'{}^2 \left(r^2 \gamma ' f'-6\right)+f^2 \bigg(f' \left(5 r^3 \gamma '{}^3-20 r \gamma '-16\right) \nn\\
%%%%%%%%%%
&\quad& +2 \gamma ' \left(r \gamma '-8\right)\bigg)\bigg]\Big/(2 r^5) +8 \pi  \rho +\frac{r f'+f-1}{r^2}  =0 \,, \label{tov1}\\
%%%%%%%%%%
E^r{}_r &\equiv& \frac{6 f^2 \lambda \gamma ''' \left(f'-f \gamma '\right)}{r^2}-\frac{3 f \lambda  \gamma '' \left(f^2 \left(5 r^2 \gamma '{}^2-4\right)+f \left(r^2 \gamma ' f'+4\right)-2 r^2 f'{}^2\right)}{r^4}\nn\\
%%%%%%%%%%
&\quad& -\frac{6 f^3 \lambda \gamma ''{}^2}{r^2} +\lambda f'' \left[\frac{6 f^2   \gamma ''}{r^2}-\frac{3 f \left(-3 r^2 \gamma ' f'+f \left(r^2 \gamma '{}^2+4\right)-4\right)}{r^4}\right] \nn\\
%%%%%%%%%%
&\quad& -3 \lambda  \bigg[f^3 \gamma ' \left(3 r^3 \gamma '{}^3-14 r \gamma '+16\right)+f f' \left(f' \left(2 r-r^3 \left(\gamma '\right)^2\right)-4 r \gamma '+16\right) \nn\\
%%%%%%%%%%
&\quad& +f^2 \bigg(f' \left(r^3 \gamma '{}^3+4 r \gamma '-16\right)+2 \gamma ' \left(5 r \gamma '-8\right)\bigg) \nn \\
%%%%%%%%%%
&\quad& +r f'{}^2 \left(r^2 \gamma ' f'+2\right)\bigg]\big/(2 r^5) +\frac{f r \gamma '+f-1}{r^2} -8 \pi  P(\rho) =0 \,,\label{tov2}\\
%%%%%%%%%%
E^\theta{}_\theta &\equiv& \frac{3 f^2 \lambda  \gamma '''' \left(f'-f \gamma '\right)}{r} +\lambda f''' \left[\frac{3 f^2   \gamma ''}{r}+\frac{3 f \left(r^2 \gamma ' f'-2 f+2\right)}{r^3}\right] +\frac{3 f \lambda  \gamma ' f''^2}{r} \nn \\
%%%%%%%%%%
&\quad& +\gamma ''' \left[-\frac{9 f^3 \lambda  \gamma ''}{r}+\frac{6 f^2 \lambda  f''}{r}-\frac{3 f \lambda  \left(f^2 \left(3 r^2 \left(\gamma '\right)^2-2\right)+f \left(3 r^2 \gamma ' f'+2\right)-3 r^2 \left(f'\right)^2\right)}{r^3}\right] \nn \\
%%%%%%%%%%
&\quad& +f'' \Bigg[\frac{\lambda  \left(-6 f^2 \left(r^3 \gamma '{}^3+r \gamma '-8\right)+f \left(f' \left(9 r^3 \gamma '{}^2-30 r\right)+6 \left(r \gamma '-8\right)\right)+3 r f' \left(r^2 \gamma ' f'+2\right)\right)}{2 r^4} \nn \\
%%%%%%%%%%
&\quad&  -\frac{3 f \lambda  \gamma '' \left(f \gamma '-11 f'\right)}{2 r}\Bigg] -\frac{3 f^2 \lambda  \left(\gamma ''\right)^2 \left(7 f'+11 f \gamma '\right)}{2 r} +\frac{3 \lambda \gamma '' }{4 r^4} \Big[ f^3 \left(-13 r^3 \gamma '{}^3+28 r \gamma '-32\right) \nn \\
%%%%%%%%%%
&\quad& +3 r^3 f'{}^3+f r f' \left(r^2 \gamma ' f'-12\right)+f^2 \left(f' \left(20 r-39 r^3 \gamma '{}^2\right)-20 r \gamma '+32\right) \Big] + \frac{f \gamma ''}{2} \nn \\
%%%%%%%%%%
&\quad& -\frac{3 \lambda }{4 r^5}  \Bigg[f^3 \gamma ' \left(r^4 \gamma '{}^4-6 r^2 \gamma '{}^2+28 r \gamma '-48\right)+r f'{}^2 \left(f' \left(2 r-r^3 \gamma '{}^2\right)+12\right)  \nn \\
%%%%%%%%%%
&\quad& +f f' \Big(r f' \left(5 r^3 \gamma '{}^3+2 r \gamma '-36\right)+12 \left(r^2 \gamma '^2-2 r \gamma '-4\right)\Big) \nn \\
%%%%%%%%%%
&\quad& +f^2 \Big(f' \left(7 r^4 \gamma '{}^4-30 r^2 \gamma '{}^2+40 r \gamma '+48\right)+4 \gamma ' \left(r^2 \gamma '{}^2-5 r \gamma '+12\right)\Big)\Bigg] \nn\\ 
%%%%%%%%%%
&\quad& +\frac{\left(f'+f \gamma '\right) \left(r \gamma '+2\right)}{4 r} -8 \pi  P(\rho) =0 \,,\label{tov3}
\eea
\end{subequations} 
\end{widetext}
where a prime denotes the derivative with respect to the radial coordinate $r$.
In the limit $\lambda \to 0$, one can explicitly verify that the field equations reduce to those of GR.

\section{Perturbation equations} \label{pertNSinQTG}

For completeness, we provide the explicit expressions for the non-vanishing components of the radial perturbation equations below:
\begin{widetext}
\begin{subequations} \label{radialpert0}
\bea
\delta E^t{}_t &\equiv&  \delta f \Big[2 f r^3 + \lambda  \Big[3 f \Big[4 f \left(-2 r \left(2 \gamma ''+f'' \left(r^2 \gamma ''+2\right)\right)+r \gamma '{}^2 \left(2 r^2 f''-1\right)+\gamma ' \left(8-2 r^3 f''' \right)\right) \nn\\
%%%%%%%%%%
&\quad& -3 f^2 \left(16 \gamma '+r^3 \gamma '{}^4+\gamma '{}^2 \left(8 r^3 \gamma ''-2 r\right)-8 r \gamma ''\right)+4 r f'' \left(r^2 f''+2\right) +f'{}^2 (7 r^3 \gamma '{}^2 \nn\\
%%%%%%%%%%
&\quad& -2 r (r^2 \gamma ''+7))-2 f' \Big(r \gamma ' \left(5 r^2 f''+6\right)+f \left(5 r^3 \gamma '{}^3-10 r \gamma ' \left(r^2 \gamma ''+2\right)-16\right) \nn\\
%%%%%%%%%%
&\quad& -2 r^3 f'''+8\Big)\Big] +6 e^{-\gamma } r^3 \omega ^2 \left(-f \left(2 f''+f \left(\gamma '{}^2-2 \gamma ''\right)\right)+2 f \gamma ' f'+f'{}^2\right) \Big]\Big]  \nn\\
%%%%%%%%%%
&\quad& +\delta f' \Big[\lambda  \Big[12 e^{-\gamma } f r^3 \omega ^2 \left(f \gamma '-f'\right)-3 f \Big[f^2 \left(5 r^3 \gamma '{}^3-10 r \gamma ' \left(r^2 \gamma ''+2\right)-16\right) \nn\\
%%%%%%%%%%
&\quad& +2 f \left(r \gamma ' \left(5 r^2 f''+6\right)+f' \left(2 r \left(r^2 \gamma ''+7\right)-7 r^3 \gamma '{}^2\right)-2 r^3 f'''+8\right) \nn\\
%%%%%%%%%%
&\quad& +r f' \left(3 r^2 \gamma ' f'-4 \left(r^2 f''+3\right)\right)\Big]\Big]+2 f r^4\Big] +12 f^2 \lambda  r^3 \delta f''' \left(f'-f \gamma '\right) \nn\\
%%%%%%%%%%
&\quad&  -6 f^2 \lambda  r \delta \gamma '' \left[4 f^2 \left(r^2 \gamma '{}^2-1\right)+r^2 f'{}^2+f \left(2 r^2 f''-5 r^2 \gamma ' f'+4\right)\right] +16 \pi  \delta \rho f r^5  \nn\\
%%%%%%%%%%
&\quad& +6 f \lambda  r \delta f'' \left[2 f^2 \left(-r^2 \gamma ''+r^2 \gamma '{}^2-2\right)+r^2 f'{}^2+f \left(4 r^2 f''-5 r^2 \gamma ' f'+4\right)\right]  \nn\\
%%%%%%%%%%
&\quad& -3 f \lambda  \Big[4 f^3 \left(r^3 \gamma '{}^3+r \gamma ' \left(4 r^2 \gamma ''-1\right)+4\right)+r^3 f'{}^3 -2 f r f' \left(-5 r^2 f''+7 r^2 \gamma ' f'-6\right)  \nn\\
%%%%%%%%%%
&\quad& +f^2 \Big[4 \left(\gamma ' \left(r-2 r^3 f''\right)+r^3 f'''-4\right)+5 r f' \left(-2 r^2 \gamma ''+3 r^2 \gamma '{}^2-4\right)\Big] \Big] \delta \gamma'   =0 \,, \label{starobinskypert1}\\
%%%%%%%%%%
%%%%%%%%%%
\delta E^t{}_r &\equiv& 6 f \lambda  r^2 \left(f \gamma '-f'\right) \delta f'' +3 f \lambda  \delta \gamma ' \left[4 f^2 \left(r^2 \gamma '{}^2-1\right)+r^2 f'{}^2+f \left(2 r^2 f''-5 r^2 \gamma ' f'+4\right)\right] \nn\\
%%%%%%%%%%
&\quad& +3 f \lambda  \delta f' \left[-2 r^2 f''+3 r^2 \gamma ' f'+f \left(4-3 r^2 \left(\gamma '\right)^2\right)-4\right]+\delta f \Big[3 f \lambda  \gamma ' \left(3 r^2 f''+8\right) \nn\\
%%%%%%%%%%
&\quad& +3 f^2 \lambda  \gamma ' \left(r^2 \gamma ''+4 r^2 \gamma '{}^2-12\right)+6 e^{-\gamma } \lambda  r^2 \omega ^2 \left(f'-f \gamma '\right) \nn\\
%%%%%%%%%%
&\quad& -3 \lambda  f' \left(r^2 f''+f \left(r^2 \gamma ''+4 r^2 \gamma '{}^2-8\right)+4\right)-r^3\Big]+8 \pi  \xi  r^4 (P(\rho )+\rho ) =0  \,, \label{starobinskypert2} \\
%%%%%%%%%%
%%%%%%%%%%
\delta E^r{}_r &\equiv& 12 f^3 \lambda  r^3 \delta\gamma''' \left(f \gamma '-f'\right) +6 f^2 \lambda  r {\delta f}'' \left(-3 r^2 \gamma ' f'+f \left(-2 r^2 \gamma ''+r^2 \gamma '{}^2+4\right)-4\right) \nn\\
%%%%%%%%%%
&\quad& +6 f^2 \lambda  r \delta \gamma '' \left(f^2 \left(4 r^2 \gamma ''+5 r^2 \gamma '{}^2-4\right)-2 r^2 f'{}^2+f \left(-2 r^2 f''+r^2 \gamma ' f'+4\right)\right) \nn\\
%%%%%%%%%%
&\quad& +3 f \lambda  {\delta f}' \Big[f^2 \left(r^3 \gamma '{}^3-4 \left(r^3 \gamma'''+4\right)+2 r \gamma ' \left(r^2 \gamma ''+2\right)\right) +r f' \left(3 r^2 \gamma ' f'+4\right) \nn\\
%%%%%%%%%%
&\quad& -2 f \left(r \gamma ' \left(3 r^2 f''+2\right)+r f' \left(4 r^2 \gamma ''+r^2 \gamma '{}^2-2\right)-8\right)\Big] \nn\\
%%%%%%%%%%
&\quad& +{\delta f} \Big[\lambda  \Big[3 f \Big[4 f \left(-8 \gamma '+f'' \left(4 r-2 r^3 \gamma ''\right)+r \gamma '{}^2 \left(r^2 f''+5\right)+4 r \gamma ''\right) \nn\\
%%%%%%%%%%
&\quad& +3 f^2 \left(3 r^3 \gamma '{}^4+4 \gamma ' \left(r^3 \gamma'''+4\right)+4 r \gamma '' \left(r^2 \gamma ''-2\right)+2 r \gamma '{}^2 \left(5 r^2 \gamma ''-7\right)\right)-8 r f'' \nn\\
%%%%%%%%%%
&\quad& +2 f' \left(-r \gamma ' \left(3 r^2 f''+2\right)+f \left(r^3 \gamma '{}^3-4 \left(r^3 \gamma'''+4\right)+2 r \gamma ' \left(r^2 \gamma ''+2\right)\right)+8\right)  \nn\\
%%%%%%%%%%
&\quad& -r f'{}^2 \left(4 r^2 \gamma ''+r^2 \gamma '{}^2-2\right)\Big] -6 e^{-\gamma } r \omega ^2 \Big(f^2 \left(10 r^2 \gamma ''+7 r^2 \gamma '{}^2-16\right) \nn\\
%%%%%%%%%%
&\quad& -2 f \left(r^2 f''-8\right)+3 r^2 f'{}^2\Big)\Big]-2 f r^3 \left(r \gamma '+1\right)\Big] +\delta \gamma ' \Big[-2 f^2 r^4 \nn\\
%%%%%%%%%%
&\quad&  + \lambda \Big[3 f \Big(4 f^3 \left(3 r^3 \gamma '{}^3+r^3 \gamma'''+r \gamma ' \left(5 r^2 \gamma ''-7\right)+4\right) -2 f r f' \left(3 r^2 f''+r^2 \gamma ' f'+2\right) \nn\\
%%%%%%%%%%
&\quad& +r^3 f'{}^3+f^2 \left(4 r \gamma ' \left(r^2 f''+5\right)+r f' \left(2 r^2 \gamma ''+3 r^2 \gamma '{}^2+4\right)-16\right)\Big) \nn \\
%%%%%%%%%%
&\quad&  -12 e^{-\gamma } f^2 r^3 \omega ^2 \left(f \gamma '-f'\right)\Big]\Big] +16 \pi  \delta \rho f r^5 P'(\rho ) =0 \,, \label{starobinskypert3}\\
%%%%%%%%%%
%%%%%%%%%%
\delta E^\theta{}_\theta &\equiv& 12 f^4 \lambda  \left(f'-f \gamma '\right) \delta\gamma'''' r^4+12 f^3 \lambda  \left(f' \gamma ' r^2+f \left(r^2 \gamma ''-2\right)+2\right) \delta f''' r^2  \nn \\
%%%%%%%%%%
&\quad& -12 f^3 \lambda  \left[\left(3 \gamma '{}^2 r^2+3 \gamma '' r^2-2\right) f^2+\left(3 f' \gamma ' r^2-2 f'' r^2+2\right) f-3 r^2 f'{}^2\right] \delta\gamma''' r^2  \nn \\
%%%%%%%%%%
&\quad&  +\lambda {\delta f}'' \Big[12 e^{-\gamma } f^2 r^4 \omega ^2 \left(f \gamma '-f'\right)-6 f^2 r \Big(\left(2 r^3 \gamma '{}^3+r \left(\gamma '' r^2+2\right) \gamma '-4 \left(\gamma''' r^3+4\right)\right) f^2  \nn \\
%%%%%%%%%%
&\quad& -\left(2 \gamma ' \left(2 f'' r^3+r\right)+r f' \left(3 \gamma '{}^2 r^2+11 \gamma '' r^2-10\right)-16\right) f-r f' \left(f' \gamma ' r^2+2\right)\Big)\Big]  \nn \\
%%%%%%%%%%
&\quad&  +\delta \gamma '' \Big[2 f^3 r^5+\lambda  \Big(12 e^{-\gamma } f^3 r^4 \omega ^2 \left(f \gamma '-f'\right)-3 f^2 r \Big[\Big(13 r^3 \gamma '{}^3+4 r \left(11 r^2 \gamma ''-7\right) \gamma '   \nn \\
%%%%%%%%%%
&\quad& +4 \left(3 \gamma''' r^3+8\right)\Big) f^3+\Big(2 r \gamma ' \left(f'' r^2+10\right)+r f' \left(39 \gamma '{}^2 r^2+28 \gamma '' r^2-20\right)  \nn \\
%%%%%%%%%%
&\quad& -4 \left(f''' r^3+8\right)\Big) f^2-r f' \left(f' \gamma ' r^2+22 f'' r^2-12\right) f-3 r^3 f'{}^3\Big]\Big)\Big] -32 f^2 \pi \delta \rho P'(\rho ) r^5 \nn \\
%%%%%%%%%%
&\quad&  +\delta \gamma ' \Big[f^2 \left(r f'+2 f \left(r \gamma '+1\right)\right) r^4+\lambda  \Big(6 e^{-\gamma } f^2 r^2 \omega ^2 \Big(\left(3 \gamma '{}^2 r^2+2 \gamma '' r^2-4\right) f^2+4 f  \nn \\
%%%%%%%%%%
&\quad& -r^2 f'{}^2\Big)-3 f^2 \Big[-2 f'^2 \left(f' \gamma '+f''\right) r^4+f \Big(r \left(15 \gamma '{}^2 r^2-\gamma '' r^2+2\right) f'^2  \nn \\
%%%%%%%%%%
&\quad& -4 \left(f''' r^3+3 \gamma ' \left(r^2 f''-2\right) r+6\right) f'-4 r f'' \left(f'' r^2+1\right)\Big) r+2 f^2 \Big(f'' \gamma '' r^4  \nn \\
%%%%%%%%%%
&\quad& +2 f'' r^2+10 \gamma '' r^2-20 \gamma ' r+f' \left(14 \gamma '{}^3 r^3+6 \gamma''' r^3+\gamma ' \left(39 r^3 \gamma ''-30 r\right)+20\right) r  \nn \\
%%%%%%%%%%
&\quad& +6 \gamma '{}^2 \left(f'' r^4+r^2\right)+24\Big)+f^3 \Big(5 \gamma '{}^4 r^4+22 \gamma ''{}^2 r^4+4 \gamma'''' r^4-28 \gamma '' r^2  \nn \\
%%%%%%%%%%
&\quad&  +3 \gamma '{}^2 \left(13 r^2 \gamma ''-6\right) r^2+8 \gamma ' \left(3 \gamma''' r^3+7\right) r-48\Big)\Big]\Big)\Big] +{\delta f}' \Big[f^2 \left(r \gamma '+2\right) r^4 \nn \\
%%%%%%%%%%
&\quad&  +\lambda  \Big[3 \Big(-\Big(7 \gamma '{}^4 r^4+14 \gamma ''{}^2 r^4-4 \gamma'''' r^4-20 \gamma '' r^2+4 \gamma ' \left(3 \gamma''' r^3+10\right) r  \nn \\
%%%%%%%%%%
&\quad&  +\gamma '{}^2 \left(39 r^4 \gamma ''-30 r^2\right)+48\Big) f^2-2 \Big(10 f'' r^2-11 f'' \gamma '' r^4+6 \gamma '' r^2  \nn \\
%%%%%%%%%%
&\quad&  -2 \gamma ' \left(f''' r^3+6\right) r+f' \left(5 r^3 \gamma '{}^3+\left(2 r-r^3 \gamma ''\right) \gamma '-12 \left(\gamma''' r^3+3\right)\right) r \nn \\
%%%%%%%%%%
&\quad& +\gamma '{}^2 \left(6 r^2-3 r^4 f''\right)-24\Big) f+4 r^2 f''+4 r f' \left(r^3 \gamma ' f''-6\right) \nn \\
%%%%%%%%%%
&\quad& +3 r^2 f'{}^2 \left(\gamma '{}^2 r^2+3 \gamma '' r^2-2\right)\Big) f^2+6 e^{-\gamma } r^2 \omega ^2 \Big(\left(3 \left(\gamma '\right)^2 r^2+10 \gamma '' r^2-12\right) f^2  \nn \\
%%%%%%%%%%
&\quad& +4 \left(f' \gamma ' r^2-f'' r^2+3\right) f+3 r^2 \left(f'\right)^2\Big) f\Big]\Big]  +{\delta f} \Big[f^2 \left(r \gamma '^2+2 \gamma '+2 r \gamma ''\right) r^4 \nn \\
%%%%%%%%%%
&\quad&  -2 e^{-\gamma } f \omega ^2 r^5 +\lambda  \Big[-12 e^{-2 \gamma } f r^4 \left(f \gamma '-f'\right) \omega ^4 +6 e^{-\gamma } r \Big(\Big(\gamma '{}^3 r^3+10 \gamma''' r^3  \nn \\
%%%%%%%%%%
&\quad&  +\gamma ' \left(9 r^2 \gamma ''-4\right) r+32\Big) f^3+\left(5 \gamma ' f'' r^3-2 f''' r^3+f' \left(2 \gamma '{}^2 r^2+9 \gamma '' r^2-8\right) r-32\right) f^2  \nn \\
%%%%%%%%%%
&\quad&  +r f' \left(-2 f' \gamma ' r^2+5 f'' r^2-4\right) f-3 r^3 \left(f'\right)^3\Big) \omega ^2+3 f^2 \Big(4 \gamma ' f''{}^2 r^4+4 \gamma ' f'' r^2  \nn \\
%%%%%%%%%%
&\quad&  +8 f''' r^2-32 f'' r+f'{}^2 \left(-5 r^3 \gamma '{}^3+r \left(r^2 \gamma ''-2\right) \gamma '+12 \gamma''' r^3+3\right)\Big) r   \nn \\
%%%%%%%%%%
&\quad&  +2 f' \left(3 \gamma '{}^2 \left(r^2 f''-2\right) r^2-6 \gamma '' r^2+f'' \left(11 r^2 \gamma ''-10\right) r^2+2 \gamma ' \left(f''' r^3+6\right) r+24\right)  \nn \\
%%%%%%%%%%
&\quad&  -3 f^2 \Big(r^4 \gamma '{}^5+r^2 \left(13 r^2 \gamma ''-6\right) \gamma '{}^3+4 r \left(3 \gamma''' r^3+7\right) \gamma '{}^2+(22 \gamma ''{}^2 r^4+4 \gamma'''' r^4  \nn \\
%%%%%%%%%%
&\quad& -28 \gamma '' r^2-48) \gamma '+4 r \left(\gamma '' \left(3 \gamma''' r^3+8\right)-2 r \gamma'''\right)\Big) -2 f \Big(4 \left(f'' r^4+r^2\right) \gamma '{}^3  \nn \\
%%%%%%%%%%
&\quad& -20 r \gamma '{}^2+2 \left(10 \gamma '' r^2+f'' \left(\gamma '' r^2+2\right) r^2+24\right) \gamma '-4 r \Big(\gamma '' \left(f''' r^3+8\right) \nn \\
%%%%%%%%%%
&\quad& -2 r \left(f'''+\gamma'''\right)+2 f'' \left(\gamma''' r^3+4\right)\Big)+f' \Big(7 \gamma '{}^4 r^4+14 \gamma ''{}^2 r^4-4 \gamma'''' r^4-20 \gamma '' r^2 \nn \\
%%%%%%%%%%
&\quad& +4 \gamma ' \left(3 \gamma''' r^3+10\right) r+\gamma '{}^2 \left(39 r^4 \gamma ''-30 r^2\right)+48\Big)\Big)\Big)\Big]\Big]  =0 \,, \label{starobinskypert4}
\eea
\end{subequations}
\end{widetext}
where a prime denotes the derivative with respect to the radial coordinate $r$. In the limit $\lambda \to 0$, one can explicitly verify that the perturbation equations reduce to those of GR.

\begin{acknowledgments}
We are grateful to the referee for the constructive and insightful suggestions on our paper.
We are grateful to Yong Gao, Ziyi Li, Zhonghai Liu, H. L\"u, Lijing Shao, Rui Xu, and Fangkang Yu, for useful discussions. We are particularly indebted to the referee for their many constructive and valuable comments.
S.L. and H.Y. were supported in part by the National Natural Science Foundation of China (No. 12105098, No. 12481540179, No. 12075084, No. 11690034, No. 11947216, and No. 12005059) and the Natural Science Foundation of Hunan Province (No. 2022JJ40264), and the innovative research group of Hunan Province under Grant No. 2024JJ1006, and by the Excellent Young Scholars Program of the Hunan Provincial Department of Education under Grant No. 25B0092. 

\end{acknowledgments}

\section*{Data Availability Statement}

There are no publicly available research data or software supporting this manuscript. Requests for further information or data should be sent to the authors.


\end{document}